\documentclass[sigconf]{acmart}

\AtBeginDocument{%
  }

\copyrightyear{2026}
\acmYear{2026}
\setcopyright{cc}
\setcctype{by}
\acmConference[KDD '26]{Proceedings of the 32nd ACM SIGKDD Conference on Knowledge Discovery and Data Mining V.2}{August 09--13, 2026}{Jeju Island, Republic of Korea}
\acmBooktitle{Proceedings of the 32nd ACM SIGKDD Conference on Knowledge Discovery and Data Mining V.2 (KDD '26), August 09--13, 2026, Jeju Island, Republic of Korea}
\acmDOI{10.1145/3770855.3818978}
\acmISBN{979-8-4007-2259-2/2026/08}

\usepackage{algorithm}
\usepackage{algorithmic}
\usepackage{multirow}
\usepackage{makecell}
\usepackage{subcaption}
\usepackage[nameinlink]{cleveref}

\newif\ifFMAllowMissingFigures
\FMAllowMissingFiguresfalse

\ifFMAllowMissingFigures
  \makeatletter
  \let\FM@orig@includegraphics\includegraphics
  \renewcommand{\includegraphics}[2][]{%
    \IfFileExists{#2}{%
      \FM@orig@includegraphics[#1]{#2}%
    }{%
      \begingroup
        \setlength{\fboxsep}{2pt}%
        \fbox{\scriptsize Missing figure: \texttt{\detokenize{#2}}}%
      \endgroup
    }%
  }
  \makeatother
\fi

\begin{document}
\raggedbottom

\title{FactorMiner: A Self-Evolving Agent with Skills and Experience Memory for Financial Alpha Discovery}

\author{Yanlong Wang}
\affiliation{%
  \institution{Tsinghua Shenzhen International Graduate School, Tsinghua University}
  \institution{Pengcheng Laboratory}
  \city{Shenzhen}
  \country{China}}
\email{wangyanl21@mails.tsinghua.edu.cn}

\author{Jian Xu}
\affiliation{%
  \institution{Tsinghua Shenzhen International Graduate School, Tsinghua University}
  \city{Shenzhen}
  \country{China}}
\email{xujian20@mails.tsinghua.edu.cn}

\author{Hongkang Zhang}
\affiliation{%
  \institution{Tsinghua Shenzhen International Graduate School, Tsinghua University}
  \city{Shenzhen}
  \country{China}}
\email{zhanghk21@mails.tsinghua.edu.cn}

\author{Shao-Lun Huang}
\authornote{Corresponding authors.}
\affiliation{%
  \institution{Tsinghua Shenzhen International Graduate School, Tsinghua University}
  \city{Shenzhen}
  \country{China}}
\email{shaolun.huang@sz.tsinghua.edu.cn}

\author{Danny Dongning Sun}
\authornotemark[1]
\affiliation{%
  \institution{Pengcheng Laboratory}
  \institution{Shenzhen Finance Institute, Chinese University of Hong Kong}
  \city{Shenzhen}
  \country{China}}
\email{ds316@columbia.edu}

\author{Xiao-Ping Zhang}
\authornotemark[1]
\affiliation{%
  \institution{Tsinghua Shenzhen International Graduate School, Tsinghua University}
  \city{Shenzhen}
  \country{China}}
\email{xpzhang@ieee.org}

\renewcommand{\shortauthors}{Yanlong Wang et al.}

\begin{abstract}
  Formulaic alpha factor mining is a critical yet challenging task in quantitative investment, characterized by a vast search space and the need for domain-informed, interpretable signals. However, finding novel signals becomes increasingly difficult as the library grows due to high redundancy. We propose \textbf{FactorMiner}, a lightweight and flexible self-evolving agent framework designed to navigate this complex landscape through continuous knowledge accumulation. FactorMiner combines a \textbf{Modular Skill Architecture} that encapsulates systematic financial evaluation into executable tools with a structured \textbf{Experience Memory} that distills historical mining trials into actionable insights (successful patterns and failure constraints). By instantiating the Ralph Loop paradigm---retrieve, generate, evaluate, and distill---FactorMiner iteratively uses memory priors to guide exploration, reducing redundant search while focusing on promising directions. Experiments on multiple datasets across different assets and markets show that FactorMiner constructs a diverse library of high-quality factors with competitive performance, while maintaining low redundancy among factors as the library scales. Overall, FactorMiner provides a practical approach to scalable discovery of interpretable formulaic alpha factors under the "Correlation Red Sea" constraint.
\end{abstract}


\maketitle

\section{Introduction}

Discovering predictive alpha factors is central to quantitative trading and portfolio construction.
In practice, automated factor discovery faces three fundamental challenges:
(i) \emph{vast search complexity}---the space of formulaic expressions grows combinatorially with operator compositions and parameters;
(ii) \emph{poor knowledge accumulation}---traditional search methods (genetic programming, reinforcement learning) fail to retain and reuse insights across exploration sessions, leading to repetitive trials;
(iii) \emph{interpretability constraints}---unlike black-box neural predictors, financial practitioners require transparent, auditable formulas with explicit financial logic for regulatory compliance and risk management.

Traditional approaches rely heavily on domain experts who manually craft factors based on financial intuition~\citep{kakushadze2016alphas,fama1993common}.
More recently, machine learning methods have been applied to asset pricing~\citep{gu2020mlassetpricing,feng2020taming}, demonstrating strong predictive power but often sacrificing interpretability.
Genetic programming~\citep{koza1992genetic} and reinforcement learning~\citep{yu2023alphagen,zhao2025quantfactor} offer automated search but suffer from knowledge forgetting: as the factor library grows and correlation constraints tighten, these methods lack mechanisms to accumulate structural knowledge about what works and what fails.
Moreover, existing methods treat each mined factor in isolation without considering the global factor library perspective, they optimize individual factor quality but ignore how new factors interact with the existing library, leading to redundant discoveries and inefficient exploration.

We study formulaic factor mining as a task for self-evolving AI agents~\citep{yao2023react,shinn2023reflexion}.
Each factor is an explicit expression over market fields (e.g., close, volume, VWAP) composed from a library of 60+ operators.
Unlike end-to-end neural approaches, formulaic factors enable human-in-the-loop auditing and compositional generalization across market regimes.
The key research question is: \textit{how can an agent efficiently explore this vast program space while maintaining a global view of the factor library and autonomously accumulating structural knowledge across exploration sessions?}

We propose \textbf{FactorMiner}, a self-evolving agent framework that addresses these challenges through two synergistic mechanisms: a compositional skill architecture and experience memory.
First, we design factor mining as a reusable agent skill that can be invoked on-demand by a language model agent~\citep{schick2023toolformer,yao2023react}.
The skill encapsulates domain knowledge---a curated operator library with 60+ financial operators, a multi-stage validation pipeline with IC thresholds and correlation checks, and standardized evaluation protocols---enabling flexible task decomposition and independent upgrades without retraining the agent.
Second, we introduce experience memory that enables the agent to self-evolve through accumulated knowledge~\citep{shinn2023reflexion,park2023generativeagents}.
The memory stores distilled structural patterns from historical mining sessions: successful patterns (factor templates that consistently pass quality thresholds) and forbidden regions (factor families with high mutual correlation to existing library members).
Crucially, this memory maintains a global factor library perspective: the agent decides mining directions by considering how candidate factors complement the existing library, rather than optimizing individual factors in isolation.

The agent adopts the Ralph Loop paradigm for self-evolution: retrieve relevant patterns from experience memory, generate candidate factors by invoking the mining skill with retrieved priors, evaluate candidates through parallel validation, and distill outcomes back into memory.
This creates a positive feedback cycle where each mining session improves future exploration efficiency, enabling the agent to continuously refine its search strategy.

\begin{figure*}[htbp]
\centering
\includegraphics[width=0.95\textwidth]{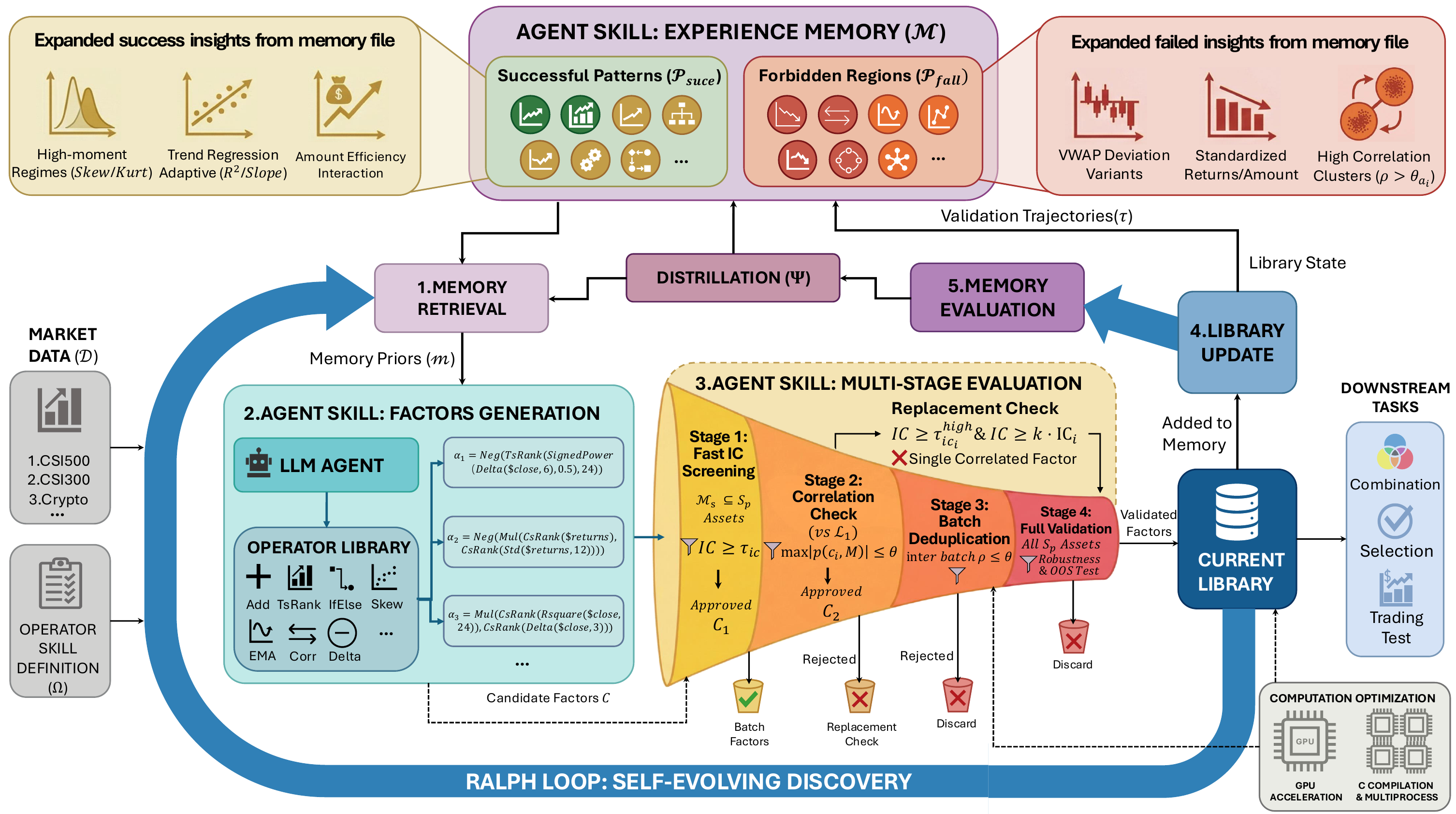}
\caption{FactorMiner System Architecture. The Ralph Loop framework integrates three key components: (1) Experience Memory that stores successful patterns and forbidden regions from past mining sessions; (2) Agent Skill that encapsulates the multi-stage validation pipeline (IC screening, correlation checking, deduplication, and full validation); (3) Factor Library that grows dynamically while maintaining orthogonality constraints. The agent iteratively retrieves memory priors, generates candidates through the skill, and distills outcomes back into memory for improved future exploration.}
\label{fig:main-process}
\Description{}
\end{figure*}

To ensure computational efficiency, we build a lightweight yet high-performance system leveraging GPU-accelerated operators, multi-process parallelization, and C-compiled efficient numerical operations. This yields substantial speedups over standard Python implementations (e.g., NumPy/Pandas), making large-scale iterative evaluation computationally feasible. We summarize our main contributions as follows:
{\setlength{\leftmargini}{1.4em}%
\begin{itemize}
  \setlength{\topsep}{2pt}
  \setlength{\itemsep}{1pt}
  \setlength{\parsep}{0pt}
  \item \textbf{Experience memory for agent-based factor mining.} We introduce structured experience memory into agent-based factor discovery, enabling self-evolution through accumulated knowledge. The memory stores reusable patterns that guide exploration and reduce redundant search compared to memoryless baselines.
  
  \item \textbf{Modular skill architecture for on-demand invocation.} We design factor mining as a compositional, reusable skill that encapsulates domain knowledge in a standalone module. This enables flexible agent invocation, independent skill upgrades, and clear separation between agent reasoning and skill execution.
  
  \item \textbf{Lightweight and efficient mining system.} We build a high-performance factor evaluation engine using GPU accelerated operators, multi-process parallelization, and C-compiled efficient numerical operations, enabling large-scale factor mining. This supports rapid iteration and large-scale factor exploration while adapting to different server configurations and compute budgets.
  
  \item \textbf{Global factor library perspective.} We incorporate factor library admission mechanisms into the mining loop, enabling the agent to make mining decisions from a global library perspective. The agent considers how candidate factors complement the existing library, rather than optimizing individual factors in isolation.
  
  \item \textbf{Open factor library with research and practical value.} We provide 110 A-share equity factors with explicit formulaic expressions, validated on real market data. These interpretable factors serve as diagnostic tools to understand cross-market behavioral anomalies and market microstructure inefficiencies, offering both research insights and practical value for quantitative trading.
\end{itemize}
}

\section{Related Work}

\subsection{Automated Alpha Factor Discovery}

Formulaic alpha libraries provide explicit, human-readable expressions valuable for interpretable research. \citet{kakushadze2016alphas} presents 101 formulaic alphas with explicit expressions and low mutual correlations (15.9\%), showing that compact formula libraries can be both diverse and effective; widely used industry variants such as Alpha191 extend this with around 191 expressions covering additional microstructure patterns. The discovery process behind such libraries is often undocumented and may combine expert design with automated search (e.g., genetic programming), but the released artifacts are explicit formulas. These classical libraries remain largely limited to daily-frequency data, and the interpretability of deeply nested expressions can degrade, so high-frequency markets still lack comparable curated factor libraries for systematic research.

To automate factor discovery, evolutionary program search methods such as genetic programming represent factor formulas as executable programs and evolve them via crossover and mutation~\citep{koza1992genetic,neely1997technical,allen1999genetic}. In practice, vanilla genetic programming can explore the expression space inefficiently, exhibiting slow convergence and limited semantic guidance because genetic operators primarily act on program syntax rather than program behavior~\citep{poli2008fieldguide,moraglio2012geometric}. More recently, machine learning methods have been applied to empirical asset pricing and high-dimensional model selection~\citep{gu2020mlassetpricing,kozak2020shrinking,feng2020taming,harvey2016crosssection,harvey2021lucky}. While these methods achieve strong predictive performance, they are often less transparent than formulaic signals and can be difficult to interpret or audit in high-stakes settings~\citep{rudin2019stop}.

Reinforcement learning has been used to navigate the discrete space of formulaic factor expressions by treating evaluation metrics (e.g., IC/ICIR) as rewards~\citep{yu2023alphagen,zhao2025quantfactor}, often at the cost of additional training and repeated evaluation overhead. In parallel, neural and {LLM}-driven frameworks generate and refine formulaic factors with exploration strategies designed to mitigate decay and redundancy~\citep{shi2024alphaforge,tang2025alphaagent}. Despite these advances, these approaches still face the challenge of explicitly and persistently reusing structural patterns across mining sessions; as the factor library grows and redundancy/correlation constraints tighten, exploration can become increasingly repetitive~\citep{feng2020taming,kozak2020shrinking}.

\subsection{AI Agents with Skills and Memory}

Recent language-model agents shift from one-shot text generation to closed-loop task execution via tool calls and feedback. Toolformer~\citep{schick2023toolformer} trains LLMs to insert API calls in a self-supervised manner, while ReAct~\citep{yao2023react} interleaves reasoning traces with actions---the agent plans, invokes tools, observes outcomes, and iterates. This tool-augmented, skill-based view motivates our design: factor mining is packaged as an executable skill that an agent can invoke on demand. Packaging domain procedures as reusable skills lets an agent tackle complex, specialized tasks without fine-tuning model weights, while cutting repeated in-context instructions and their token overhead and improving reliability on multi-step domain workflows.

Memory and self-improvement mechanisms further let agents accumulate experience over time. Reflexion~\citep{shinn2023reflexion} uses language-based self-reflection to store lessons in natural language, and generative agents~\citep{park2023generativeagents} maintain memory streams (e.g., observations, reflections, plans) to condition future behavior; more broadly, continual learning studies how to retain knowledge without catastrophic forgetting~\citep{parisi2019continual}. In our setting, we instantiate memory for symbolic program synthesis: instead of storing dense representations or replay buffers~\citep{blundell2016model}, we retain reusable symbolic rules and structural patterns discovered during factor mining, together with summary statistics that help avoid redundant regions of the search space.

From a theoretical perspective, meta-learning and meta-RL formalize "learning to learn" by extracting transferable learning biases from prior tasks or episodes~\citep{finn2017maml,nichol2018reptile,gupta2018meta}. Analogously, a memory-guided mining process can be seen as acquiring search priors: beyond producing individual factors, the system accumulates structural knowledge that shapes future exploration.

Beyond learning to call tools, systems work has advocated modular agent architectures that route subproblems to specialized models, external knowledge sources, and discrete reasoning modules (e.g., MRKL~\citep{karpas2022mrkl}; HuggingGPT~\citep{shen2023hugginggpt}). In parallel, recent benchmarks and training pipelines study how to ground LLM outputs into executable API calls at scale and reduce tool hallucination, including API-Bank~\citep{li2023apibank}, ToolLLM/ToolBench~\citep{qin2023toollm}, and Gorilla/APIBench~\citep{patil2023gorilla}. These works collectively emphasize the importance of separating high-level planning from reliable execution, and of representing tools/skills in a form that supports retrieval and continual updates.

Another line of research addresses long-horizon agent behavior under limited context windows by introducing explicit long-term memory managers. MemGPT~\citep{packer2024memgpt} proposes OS-inspired hierarchical memory with controlled paging between fast and slow memory, while Voyager~\citep{wang2023voyager} demonstrates an open-ended embodied agent that accumulates an executable skill library and reuses it to generalize to new tasks. Such settings are often effectively non-stationary: as an agent acquires new tools/skills (or as an external library grows), the feasible action space and redundancy patterns evolve, motivating experience summarization mechanisms that capture reusable patterns and avoid repeatedly exploring known dead ends.

\section{Methodology}

\subsection{Problem Formulation}

We define the alpha factor discovery task over a universe of $M$ assets and a time horizon $T$. 
The raw market input is represented by a tensor $\mathcal{D} \in \mathbb{R}^{M \times T \times F}$, where each entry $d_{m,t}^{(j)}$ denotes the value of feature $j$ (e.g., price, volume) for asset $m$ at time $t$. 
\noindent\textbf{Symbolic factor space.}
A symbolic alpha factor $\alpha$ is a computational program composed from an operator set $\Omega$ that transforms market states into a cross-sectional predictive signal $\mathbf{s}_t \in \mathbb{R}^M$:
\begin{equation}
  \alpha: \mathcal{D}_{:, :t, :} \rightarrow \mathbf{s}_t, \quad s_{t}^{(m)} = \alpha(\mathbf{d}_{m,:t})
\end{equation}
where $s_{t}^{(m)}$ represents the predictive score for asset $m$ relative to its peers.
Each operator $o \in \Omega$ has a fixed arity and a typed signature (e.g., time-series $\rightarrow$ time-series, cross-section $\rightarrow$ cross-section), and programs $\alpha \in \mathcal{P}(\Omega)$ correspond to expression trees composed from $\Omega$ with admissible parameters; detailed definitions and categories of $\Omega$ are provided in Appendix~\ref{app:factor-details} and Table~\ref{tab:operators}.

The effectiveness of $\alpha$ is quantified by the Information Coefficient (IC), defined as the cross-sectional Spearman rank correlation between the signal $\mathbf{s}_t$ and subsequent returns $\mathbf{r}_{t+1} \in \mathbb{R}^M$:
\begin{equation}
  \text{IC}_t(\alpha) = \text{Corr}_{\text{rank}}\bigl(\mathbf{s}_t, \mathbf{r}_{t+1}\bigr)
\end{equation}
Consistency over time is measured by the Information Ratio (ICIR): $\text{ICIR}(\alpha) = \mu(\text{IC}_t) / \sigma(\text{IC}_t)$.

\noindent\textbf{Correlation metric.}
To enforce library diversity, we measure redundancy between two factors $\alpha$ and $\beta$ by the time-average cross-sectional Spearman correlation of their realized signals:
\begin{equation}
\label{eq:rho}
  \rho(\alpha,\beta) = \frac{1}{|\mathcal{T}|}\sum_{t \in \mathcal{T}} \text{Corr}_{\text{rank}}\bigl(\mathbf{s}_t(\alpha), \mathbf{s}_t(\beta)\bigr),
\end{equation}
where $\mathcal{T}$ denotes the set of evaluation timestamps. This definition matches our current implementation, but the framework can accommodate alternative dependence measures (e.g., time-series correlation, partial correlation, or non-linear dependence metrics) by replacing $\rho(\cdot,\cdot)$ accordingly.

\textbf{Objective: Orthogonal Library Synthesis.}
FactorMiner treats the problem as the iterative construction of a diverse factor library $\mathcal{L} = \{\alpha_1, \ldots, \alpha_K\}$. 
The goal is to maximize the aggregate predictive quality of the library subject to a global redundancy constraint:
\begin{equation}
\label{eq:objective}
  \mathcal{L}^* = \arg\max_{\mathcal{L} \subset \mathcal{P}} \sum_{\alpha \in \mathcal{L}} \Phi(\alpha) \quad \text{s.t.} \quad \forall \alpha_i \neq \alpha_j \in \mathcal{L}: |\rho(\alpha_i, \alpha_j)| < \theta
\end{equation}
where $\mathcal{P}$ is the infinite space of constructible programs, $\Phi(\cdot)$ is a fitness metric, and $\theta$ is the correlation budget.

\textbf{The Correlation Red Sea.}
As the library $\mathcal{L}$ populates, the feasible region for new orthogonal factors---$\mathcal{P}_{\text{orth}} = \{\alpha \in \mathcal{P} : \max_{g \in \mathcal{L}} |\rho(\alpha, g)| < \theta\}$---shrinks rapidly as diversity constraints tighten. 
Standard search methods (e.g., GP or RL) often get trapped in this correlation red sea because they lack mechanisms to track explored regions or failure patterns.

\textbf{Decision-Theoretic Memory Formulation.}
To navigate the correlation red sea, we reformulate the discovery process as a sequential decision task over an evolving internal knowledge state $S_t = (\mathcal{L}_t, \mathcal{M}_t)$, where $\mathcal{M}_t$ represents the persistent experience memory. 
The agent first retrieves a context-dependent memory signal $m_t$ from $\mathcal{M}_t$ and $\mathcal{L}_t$, and then samples candidates from a memory-conditioned policy:
\begin{equation}
\label{eq:conditional-policy}
\alpha \sim \pi(\alpha \mid m_t), \quad m_t = R(\mathcal{M}_t,\mathcal{L}_t)
\end{equation}
The role of $\mathcal{M}$ is to induce a probability measure contraction over the program space $\mathcal{P}$. 
By distilling historical trajectories $\tau = \{(\alpha_i, R_i)\}_{i=1}^B$ into structured patterns, $\mathcal{M}$ shifts the sampling mass toward the orthogonal manifold $\mathcal{P}_{\text{orth}}$. 
Mathematically, the evolution of memory is governed by a distillation operator $\Psi$:
\begin{equation}
\label{eq:distillation}
\mathcal{M}_{t+1} = \Psi(\mathcal{M}_t, \tau_t)
\end{equation}
which updates the agent's belief distribution so that future exploration is steered toward higher-utility and lower-redundancy regions of $\mathcal{P}$.

\subsection{Factor Mining Skill Architecture}
\label{sec:skills}

Unlike traditional RL or monolithic agent approaches where domain logic is often hard-coded or entangled with the agent's reasoning loop, FactorMiner adopts a modular skill-based architecture. 
Inspired by the tool-augmented paradigm~\citep{schick2023toolformer}, we encapsulate the entire factor mining process as a standalone, reusable Agent Skill---a standardized interface that exposes high-level capabilities to the LLM while abstracting away low-level execution details.
The overall interaction pattern is illustrated in \Cref{fig:main-process}.

\textbf{Compositional Design.} 
The skill is structured as a hierarchical library of tools:
{\setlength{\leftmargini}{1.4em}%
\begin{itemize}
    \item \textbf{Operator Layer}: A curated set of 60+ financial operators (e.g., \texttt{TsRank}, \texttt{Rsquare}) implemented with GPU-accelerated backends. This ensures that the agent's symbolic proposals are executable and computationally efficient.
    \item \textbf{Validation Pipeline}: A rigorous, standardized protocol for factor assessment (\texttt{check\_ic} $\rightarrow$ \texttt{check\_correlation} $\rightarrow$ \texttt{admit}). By decoupling validation from generation, we ensure that the agent's "creativity" is bounded by strict quantitative constraints, reducing hallucinated or invalid candidates.
\end{itemize}
}

\textbf{Advantages of the Skill-Based Approach.}
This modular design offers three distinct advantages over monolithic architectures:
\begin{enumerate}
    \item \textbf{Prevention of Calculation Hallucination}: Large language models often struggle with precise arithmetic and algorithmic execution. By offloading the evaluation to a deterministic, code-based skill, FactorMiner avoids "hallucinated metrics": the reported IC and ICIR are computed deterministically.
    \item \textbf{Cross-Domain Transferability}: The skill is parameterized to support multiple markets (e.g., A-shares vs. Crypto) via configuration files. The same high-level agent reasoning logic can be applied to different financial domains simply by switching the underlying skill context, as demonstrated in our cross-market experiments (Section~\ref{sec:results-robustness}).
    \item \textbf{Independent Optimization}: The skill's execution backends and evaluation pipeline can be optimized independently of the agent's reasoning model, improving throughput without retraining the LLM backbone.
\end{enumerate}

\subsection{Experience Memory}
\label{sec:memory}
\label{app:memory}

A key component of FactorMiner is the experience memory $\mathcal{M}$, a structured knowledge base that accumulates insights from historical mining sessions. 
We formalize the dynamics of the memory system through three conceptual operators: Formation, Evolution, and Retrieval.

\textbf{Memory Formation.}
At the end of each mining batch $t$, the agent analyzes the mining trajectory $\tau_t = \{(\alpha_i, R_i)\}_{i=1}^B$, where $R_i$ represents the evaluation feedback (IC, correlation, etc.). A formation operator $F$ selectively extracts informational artifacts:
\begin{equation}
\mathcal{M}_{t+1}^{\text{form}} = F(\mathcal{M}_t, \tau_t)
\end{equation}
This process distills raw data into symbolic patterns, categorizing them into \textbf{successful patterns} $\mathcal{P}_{\text{succ}}$ (those that pass admission) and \textbf{forbidden regions} $\mathcal{P}_{\text{fail}}$ (those rejected due to high correlation).

\textbf{Memory Evolution.}
Formed memory candidates are integrated into the existing knowledge base through an evolution operator $E$:
\begin{equation}
\mathcal{M}_{t+1} = E(\mathcal{M}_t, \mathcal{M}_{t+1}^{\text{form}})
\end{equation}
This operator consolidates redundant entries and discards low-utility information. For instance, if a specific \texttt{VWAP\_Deviation} variant is admitted but shows 0.82 correlation with existing factors, it is reclassified into $\mathcal{P}_{\text{fail}}$ to prevent further redundant exploration.

\textbf{Memory Retrieval.}
During the factor generation phase, the agent retrieves a context-dependent memory signal $m_t$ via a retrieval operator $R$:
\begin{equation}
m_t = R(\mathcal{M}_t, \mathcal{L}_t)
\end{equation}
The signal $m_t$ serves as a prompt-level constraint for the LLM policy, effectively shaping the sampling distribution $\pi(\alpha \mid m_t)$. 
In practice, we store experience as compact natural-language templates with canonical examples (e.g., "recommended directions" and "forbidden directions"), and retrieve them by matching against the current library diagnostics and recent rejection reasons.

\textbf{Memory Content.}
The resulting memory $\mathcal{M}$ maintains a persistent record of the evolving mining landscape:

\textbf{(1) Mining State ($\mathcal{S}$):} Tracks the global evolution of the factor library, including current library size $|\mathcal{L}|$, recent admission logs, and saturation metrics that signal when specific logical domains (e.g., price-volume reversal) are becoming overpopulated.

\textbf{(2) Structural Experience ($\mathcal{P}$):} This is the core of the agent's guidance system, categorized into:
{\setlength{\leftmargini}{1.4em}%
\begin{itemize}
    \item \textbf{Recommended Directions ($\mathcal{P}_{\text{succ}}$):} High-success logical templates distilled from recent batches, such as higher moment regimes (using Skew/Kurt for environment switching) and robust efficiency interaction.
    \item \textbf{Forbidden Directions ($\mathcal{P}_{\text{fail}}$):} Regions identified as "Red Seas" due to persistent high correlation with the existing library, such as simple VWAP Deviations or standardized returns.
\end{itemize}
}

\textbf{(3) Strategic Insights ($\mathcal{I}$):} High-level lessons learned from the mining process, such as the observation that non-linear combination strategies (e.g., XGBoost-based synthesis) tend to outperform linear ones, or specific operator warnings (e.g., the instability of high-order moments in high-frequency data).

\subsection{Ralph Loop: Self-Evolving Factor Discovery}

FactorMiner adopts the \textbf{Ralph Loop} paradigm---an iterative refinement philosophy where agents accumulate experience and self-evolve through repeated interaction.
We instantiate this paradigm for factor mining by integrating experience memory into the search process (Algorithm~\ref{alg:ralph}).
The key insight is that memory enables the agent to learn how to search---avoiding redundant exploration while focusing on promising regions.

\begin{algorithm}[htbp]
\caption{Ralph Loop: Self-Evolving Factor Discovery}
\label{alg:ralph}
\footnotesize
\begin{algorithmic}
\STATE \textbf{Input:} Operator library $\Omega$, experience memory $\mathcal{M}$, target library size $K$
\STATE \textbf{Output:} Factor library $\mathcal{L}$
\STATE Initialize $\mathcal{L} \leftarrow \emptyset$
\REPEAT
  \STATE \textbf{Step 1: Memory Retrieval}
  \STATE \quad Retrieve memory signal $m \leftarrow R(\mathcal{M}, \mathcal{L})$
  \STATE \textbf{Step 2: Guided Generation}
  \STATE \quad Sample batch $\mathcal{C} \sim \pi(\alpha \mid m)$ using $\Omega$
  \STATE \textbf{Step 3: Multi-Stage Evaluation}
  \STATE \quad \textbf{Stage 1}: Fast IC screening ($M_{\text{fast}}$ assets)
  \STATE \quad \quad $\mathcal{C}_1 \leftarrow \{\alpha \in \mathcal{C} : |\text{IC}(\alpha)| \geq \tau_{\text{IC}}\}$
  \STATE \quad \textbf{Stage 2}: Correlation check against $\mathcal{L}$
  \STATE \quad \quad $\mathcal{C}_2 \leftarrow \{\alpha \in \mathcal{C}_1 : \max_{g \in \mathcal{L}} |\rho(\alpha, g)| < \theta\}$
  \STATE \quad \textbf{Stage 2.5}: Replacement check
  \STATE \quad \quad For $\alpha \in \mathcal{C}_1 \setminus \mathcal{C}_2$, let $g^\star=\arg\max_{g \in \mathcal{L}} |\rho(\alpha,g)|$
  \STATE \quad \quad Replace $g^\star$ with $\alpha$ if $|\rho(\alpha,g^\star)| \ge \theta$,
  \STATE \quad \quad $\max_{g \in \mathcal{L}\setminus\{g^\star\}} |\rho(\alpha,g)| < \theta$, and $\Phi(\alpha) \ge \Phi(g^\star) + \Delta$
  \STATE \quad \textbf{Stage 3}: Batch deduplication (intra-batch $\rho < \theta$)
  \STATE \quad \textbf{Stage 4}: Full validation ($M_{\text{full}}$ assets) and trajectory $\tau$ collection
  \STATE \textbf{Step 4: Library Update}
  \STATE \quad Admit validated factors: $\mathcal{L} \leftarrow \mathcal{L} \cup \mathcal{C}_{\text{admitted}}$
  \STATE \textbf{Step 5: Memory Evolution}
  \STATE \quad Update strategy: $\mathcal{M} \leftarrow E(\mathcal{M}, F(\tau))$
\UNTIL{$|\mathcal{L}| \geq K$ or budget exhausted}
\end{algorithmic}
\end{algorithm}

Our instantiation of the Ralph Loop for factor mining has four key properties:

\textbf{Global library perspective.}
Unlike methods that optimize individual factors in isolation, our approach considers how each candidate complements the existing library $\mathcal{L}$.
The correlation constraint (Stage 2) ensures diversity, while the replacement mechanism (Stage 2.5) allows high-quality factors to replace inferior ones.

\textbf{Memory-guided exploration.}
By maintaining $\mathcal{P}_{\text{succ}}$ and $\mathcal{P}_{\text{fail}}$, the agent avoids redundant exploration of known failure regions while focusing on promising structural patterns.

\textbf{Multi-stage evaluation.}
The multi-stage pipeline balances efficiency and accuracy: Stage 1 uses a small asset subset for fast screening, Stages 2--3 enforce correlation constraints (inter-factor and intra-batch), and Stage 4 performs full validation only on surviving candidates.

\textbf{Self-evolution.}
After each iteration, the memory is updated via the evolution operator $E$ (integrating insights from $F$), creating a feedback loop where the agent continuously improves its search strategy.
This enables continual learning: knowledge accumulated in early sessions benefits later exploration.

\noindent\textbf{Trajectory semantics.}
The trajectory $\tau_t$ records, for each evaluated candidate, its formula $\alpha$, quality statistics (e.g., IC/ICIR), redundancy diagnostics (e.g., $\max_{g\in\mathcal{L}}|\rho(\alpha,g)|$), and the rejection or admission outcome (including whether a replacement was triggered). These fields are the inputs to the formation operator $F$ and support experience distillation across batches.

Detailed specifications of the operator library and admission criteria are provided in \Cref{app:factor-details}.

To illustrate the diversity of the resulting library, \Cref{fig:library_corr_heatmap} visualizes the correlation structure of the released full A-share factor library (110 admitted factors): most factor pairs are weakly to moderately correlated, with only a few localized clusters of higher dependence.

\section{Experiments}

\subsection{Experimental Setup}

\textbf{Datasets.} 
We evaluate FactorMiner across A-share equities and the cryptocurrency market to assess its discovery efficiency and cross-asset generalization. 
For A-share equities, we utilize intraday 10-minute bars of three index universes: the CSI 500 and CSI 1000 index constituents for mid-cap and broader small-/mid-cap coverage, and the CSI 300 index constituents for large-cap representation, with over 25 million data points in aggregate. 
For the Cryptocurrency market, we use 10-minute bars of 64 major assets from Binance. 
All datasets cover a training period from 2024-Q1 to 2024-Q4 and a held-out test period in 2025. The prediction target is instantiated as the non-overlapping next-step return ratio at the same 10-minute frequency, while the mining framework can accommodate alternative targets and horizons.

\textbf{Baselines and Metrics.} \label{sec:benchmarking}
We compare against six representative methods: (1) \emph{Alpha101 (Classic)} \citep{kakushadze2016alphas}, a static library of hand-crafted formulas; (2) \emph{Alpha101 (Adapted)}, with parameters tuned for high-frequency data; (3) \emph{Random Formula Exploration (RF)}, randomly sampling type-correct expression trees from $\mathcal{P}(\Omega)$ under a bounded depth/size distribution; (4) \emph{GPLearn} \citep{gplearn}, an enhanced genetic programming approach; (5) \emph{AlphaForge} \citep{shi2024alphaforge}, a neural framework for mining and dynamically combining formulaic alpha factors; and (6) \emph{AlphaAgent} \citep{tang2025alphaagent}, an LLM-driven proposal-refinement framework. For a fair comparison, we apply the same admission rules to each method and evaluate an equal-sized factor set; if a method admits fewer factors, we complete the set by selecting the remaining candidates with the best IC. We also include a \emph{No Memory} variant as an internal baseline in Section~\ref{sec:ablation}.
Performance is quantified using standard predictive metrics: \emph{Rank Information Coefficient} for forecasting precision, and its ratio (ICIR) for stability across time. Unless otherwise stated, we use IC to denote this Spearman correlation based IC. When applicable, all baselines share the same operator library $\Omega$, data fields, and evaluation/admission protocol, and are scored with a unified evaluation engine.
For methods that require LLM-based proposal generation, we use Gemini 3.0 Flash unless otherwise stated.
We do not fine-tune or update the LLM weights; self-evolution comes from retrieval-conditioned generation, deterministic skill execution, library admission, and memory distillation.

\begin{figure}[htbp]
  \centering
  \includegraphics[width=\columnwidth]{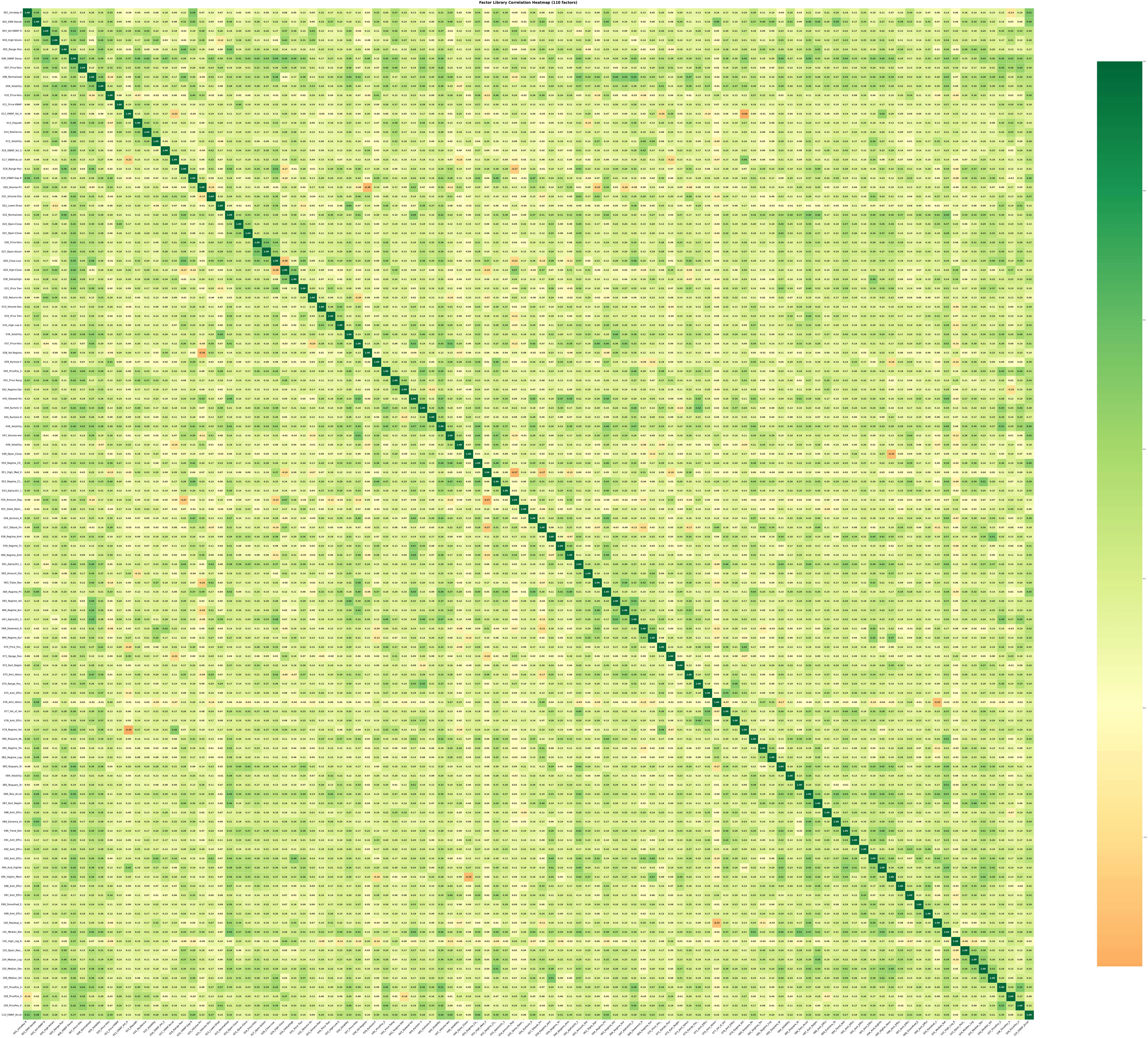}
  {\linespread{0.92}\selectfont
  \caption{Pairwise Spearman correlation heatmap of the released A-share factor library (110 admitted factors), computed from cross-sectionally standardized realized factor signals over the common time--asset panel. The average off-diagonal absolute correlation is Avg $|\rho|$ = 0.203.}\label{fig:library_corr_heatmap}\par}
  \Description{}
\end{figure}

\subsection{Main Results}
\label{sec:results}

\subsubsection{Factor Quality and Diversity}
\label{sec:results-quality}
\begin{table*}[t]
  \centering
  \caption{Out-of-Sample performance comparison with a stricter protocol. Top-40 factors are selected on CSI500 (\textbf{2024}) and evaluated on \textbf{2025} across datasets. For Alpha101, Classic is restricted to the same candidate set as Adapted. All reported IC and ICIR use the paper's absolute-IC summary: $|\mathbb{E}[IC_t]|$ and $|\mathbb{E}[IC_t]|/\mathrm{std}(IC_t)$. Factor Combination uses the frozen Top-40 with EW and ICW (weights/signs determined on 2024). Factor Selection trains on 2024 and tests on 2025 using Lasso and XGBoost.}
  \label{tab:main_results_kdd}
  \vskip 0.1in
  \scriptsize
  \setlength{\tabcolsep}{3pt}
  \begin{tabular*}{0.85\textwidth}{@{\extracolsep{\fill}}ll ccc cccc cccc}
    \toprule
    \multirow{2}{*}{\textbf{Dataset}} & \multirow{2}{*}{\textbf{Method}} & \multicolumn{3}{c}{\textbf{Factor Library (Top-40)}} & \multicolumn{4}{c}{\textbf{Factor Combination (Top-40)}} & \multicolumn{4}{c}{\textbf{Factor Selection (Train'24/Test'25)}} \\
    \cmidrule(lr){3-5} \cmidrule(lr){6-9} \cmidrule(lr){10-13}
    & & IC (\%) & ICIR & Avg $|\rho|$ & EW IC & EW ICIR & ICW IC & ICW ICIR & Las. IC & Las. ICIR & XGB. IC & XGB. ICIR \\
    \midrule
    \multirow{7}{*}{\textbf{CSI500}}
    & RF$^\ddagger$ & 2.68 & 0.25 & \textbf{0.13} & 6.98 & 0.41 & 12.02 & 0.90 & 13.81 & 1.15 & 8.99 & 0.90 \\
    & Alpha101 (Classic) & 4.49 & 0.42 & \underline{0.19} & 10.85 & 0.88 & 12.89 & 0.99 & \underline{14.09} & \textbf{1.27} & 12.07 & 1.21 \\
    & Alpha101 (Adapted) & 5.06 & 0.43 & 0.21 & \underline{11.53} & 0.86 & \underline{14.71} & \underline{1.13} & 13.70 & 1.08 & \underline{13.76} & 1.20 \\
    & GPLearn & \underline{6.04} & 0.43 & 0.44 & 10.30 & 0.62 & 13.38 & 1.00 & 12.44 & 1.17 & 9.86 & 0.95 \\
    & AlphaForge & 4.48 & 0.38 & 0.36 & 7.12 & 0.60 & 11.13 & 0.93 & 10.49 & 0.87 & 11.30 & \underline{1.25} \\
    & AlphaAgent & 5.90 & \underline{0.46} & 0.32 & 10.99 & \underline{0.92} & 11.86 & 0.99 & 13.87 & 1.19 & 11.93 & 1.24 \\
    & \textbf{FactorMiner (Ours)} & \textbf{8.25} & \textbf{0.77} & 0.31 & \textbf{14.95} & \textbf{1.29} & \textbf{15.11} & \textbf{1.31} & \textbf{14.59} & \underline{1.21} & \textbf{14.03} & \textbf{1.29} \\
    \midrule
    \multirow{7}{*}{\textbf{CSI1000}}
    & RF$^\ddagger$ & 2.88 & 0.30 & \textbf{0.13} & 7.48 & 0.49 & 12.28 & 1.00 & 13.72 & 1.24 & 9.54 & 1.03 \\
    & Alpha101 (Classic) & 4.86 & 0.50 & \underline{0.19} & 11.37 & 1.02 & 13.14 & 1.08 & \textbf{14.64} & \textbf{1.42} & 11.11 & 1.17 \\
    & Alpha101 (Adapted) & 5.32 & 0.49 & 0.21 & \underline{11.95} & 0.98 & \textbf{14.78} & \underline{1.21} & 13.08 & 1.09 & \textbf{13.88} & 1.26 \\
    & GPLearn & 5.86 & 0.48 & 0.44 & 11.10 & 0.73 & 13.66 & 1.11 & 12.87 & 1.23 & 9.53 & 0.96 \\
    & AlphaForge & 4.64 & 0.42 & 0.35 & 7.60 & 0.71 & 11.25 & 1.02 & 10.42 & 0.93 & 12.20 & \textbf{1.34} \\
    & AlphaAgent & \underline{6.21} & \underline{0.51} & 0.32 & 11.17 & \underline{1.04} & 12.00 & 1.12 & 13.73 & \underline{1.27} & 11.42 & 1.22 \\
    & \textbf{FactorMiner (Ours)} & \textbf{7.78} & \textbf{0.76} & 0.30 & \textbf{14.62} & \textbf{1.37} & \underline{14.76} & \textbf{1.39} & \underline{14.25} & 1.25 & \underline{12.42} & \underline{1.30} \\
    \midrule
    \multirow{7}{*}{\textbf{CSI300}}
    & RF$^\ddagger$ & 1.94 & 0.15 & \textbf{0.13} & 5.46 & 0.30 & 9.49 & 0.61 & 9.98 & 0.63 & 5.55 & 0.46 \\
    & Alpha101 (Classic) & 3.44 & 0.26 & \underline{0.18} & 8.55 & 0.57 & 10.31 & 0.65 & 11.84 & \underline{0.85} & 9.70 & 0.74 \\
    & Alpha101 (Adapted) & 4.00 & 0.28 & 0.20 & \underline{9.31} & \underline{0.60} & \underline{12.03} & \underline{0.76} & \underline{12.04} & 0.77 & \textbf{11.81} & 0.90 \\
    & GPLearn & 4.12 & 0.16 & 0.45 & 8.01 & 0.44 & 10.64 & 0.66 & 9.59 & 0.58 & 7.92 & 0.61 \\
    & AlphaForge & 3.53 & 0.25 & 0.36 & 5.19 & 0.35 & 8.54 & 0.57 & 6.79 & 0.43 & 10.89 & \underline{0.91} \\
    & AlphaAgent & \underline{4.69} & \underline{0.30} & 0.33 & 8.83 & 0.59 & 9.65 & 0.65 & \textbf{12.26} & \textbf{0.86} & 9.66 & 0.84 \\
    & \textbf{FactorMiner (Ours)} & \textbf{7.46} & \textbf{0.38} & 0.31 & \textbf{12.49} & \textbf{0.88} & \textbf{12.66} & \textbf{0.88} & 11.23 & 0.82 & \underline{11.33} & \textbf{0.94} \\
    \midrule
    \multirow{7}{*}{\textbf{Crypto}}
    & RF$^\ddagger$ & 1.45 & 0.09 & \textbf{0.07} & 3.31 & 0.19 & 7.91 & 0.48 & 8.04 & \underline{0.51} & 3.50 & 0.24 \\
    & Alpha101 (Classic) & 2.11 & 0.14 & \underline{0.14} & 5.91 & \underline{0.41} & 8.05 & \underline{0.53} & \textbf{9.63} & \textbf{0.54} & \underline{5.61} & \underline{0.37} \\
    & Alpha101 (Adapted) & 2.40 & 0.15 & 0.15 & 6.09 & \underline{0.41} & \underline{9.21} & \textbf{0.62} & 8.45 & 0.48 & 4.51 & 0.30 \\
    & GPLearn & 2.50 & 0.15 & 0.38 & 4.44 & 0.25 & 7.62 & 0.42 & \underline{9.07} & 0.48 & 4.04 & 0.27 \\
    & AlphaForge & 2.52 & 0.16 & 0.31 & 4.63 & 0.29 & 7.44 & 0.44 & 8.39 & 0.42 & 5.00 & 0.33 \\
    & AlphaAgent & \underline{2.86} & \underline{0.17} & 0.27 & \underline{7.94} & 0.34 & 8.40 & 0.36 & 7.48 & 0.41 & 2.35 & 0.17 \\
    & \textbf{FactorMiner (Ours)} & \textbf{3.82} & \textbf{0.28} & 0.25 & \textbf{9.48} & \textbf{0.61} & \textbf{9.48} & \textbf{0.62} & 8.98 & \underline{0.51} & \textbf{6.82} & \textbf{0.47} \\
    \bottomrule
  \end{tabular*}
  \begin{flushleft}
    \scriptsize Note: IC is computed as Spearman rank correlation per bar and summarized as $|\mathbb{E}[IC_t]|$. Las.=Lasso, XGB.=XGBoost. $^\ddagger$RF=Random Formula Exploration.
  \end{flushleft}
\end{table*}

\Cref{tab:main_results_kdd} reports 2025 out-of-sample results under a strict protocol: Top-40 factors are selected once on CSI500 (2024) and then frozen for evaluation on multiple datasets. Under this protocol, \textbf{FactorMiner} performs competitively among the compared methods across the four markets. In the Factor Library setting, it achieves IC/ICIR of \textbf{8.25\%}/\textbf{0.77} on CSI500, with similar competitive improvements on the other markets (see \Cref{tab:main_results_kdd}).

In terms of redundancy, FactorMiner's selected factor set exhibits moderate pairwise dependence. As summarized by Avg $|\rho|$ in \Cref{tab:main_results_kdd}, the average pairwise absolute correlation is 0.30--0.31 on A-shares and 0.25 on Crypto. The correlation distribution further shows a controlled tail ($|\rho|$ $\approx$ 0.44--0.45 on A-shares and 0.42 on Crypto), suggesting that the performance gains are not driven by a few near-duplicate signals. We further visualize the correlation structure of the full admitted factor library in \Cref{fig:library_corr_heatmap}. The complete factor formulas are provided in \Cref{app:factor-list}.

\subsubsection{Robustness Across Heterogeneous Markets}
\label{sec:results-robustness}
The cross-market evaluation reveals the generalization capability of the discovered factors. While the Cryptocurrency market represents a fundamentally different microstructure (24/7 trading, no price limits) and carries its own distinct risk-factor structure compared to A-shares~\citep{liu2022crypto}, FactorMiner maintains competitive performance on Crypto, with IC/ICIR of 3.82\%/0.28 in the single-factor library evaluation and 9.48\%/0.62 under a simple IC-weighted combination (see \Cref{tab:main_results_kdd}).

This robustness suggests that FactorMiner captures fundamental price-volume dynamics (e.g., liquidity-constrained reversals, volatility clustering) that are invariant across asset classes, rather than overfitting to the idiosyncrasies of a single market.

To explain why a single admission threshold need not fit all markets, we characterize each market by cross-sectional dispersion (CSD)~\citep{fei2019dispersion}, the within-bar return dispersion across assets, and return synchronicity (SYN)~\citep{morck2000synchronicity}, the degree to which assets co-move with the market aggregate (\Cref{tab:csd-syn}). A-share universes exhibit larger cross-sectional dispersion, whereas Crypto shows positive SYN, indicating stronger market-wide co-movement. This heterogeneity motivates adaptive admission: in lower-dispersion, higher-synchronicity markets such as Crypto, a lower cross-sectional IC threshold and timing-oriented objectives are more informative, while the richer cross-sectional dispersion of A-shares makes strict cross-sectional admission more meaningful.

\begin{table}[htbp]
  \centering
  \caption{Market heterogeneity measured by cross-sectional dispersion (CSD, \textperthousand) and return synchronicity (SYN) over half-year windows.}
  \label{tab:csd-syn}
  \footnotesize
  \setlength{\tabcolsep}{2pt}
  \begin{tabular*}{\columnwidth}{@{\extracolsep{\fill}}llrrrr@{}}
    \toprule
    Metric & Period & CSI500 & CSI1000 & CSI300 & Crypto \\
    \midrule
    \multirow{5}{*}{\shortstack{CSD\\(\textperthousand)}} & 2024H1 & 5.58 & 5.00 & 3.50 & 2.74 \\
    & 2024H2 & 5.05 & 10.56 & 6.44 & 2.64 \\
    & 2025H1 & 3.87 & 4.73 & 3.09 & 2.39 \\
    & 2025H2 & 4.26 & 4.41 & 3.64 & 2.48 \\
    & Avg & 4.69 & 6.18 & 4.17 & 2.56 \\
    \midrule
    \multirow{5}{*}{SYN} & 2024H1 & -1.04 & -0.80 & -1.41 & 0.47 \\
    & 2024H2 & -0.41 & -0.49 & -0.48 & 0.47 \\
    & 2025H1 & -1.31 & -1.28 & -1.40 & 0.80 \\
    & 2025H2 & -2.01 & -2.01 & -2.14 & 0.90 \\
    & Avg & -1.19 & -1.14 & -1.36 & 0.66 \\
    \bottomrule
  \end{tabular*}
\end{table}

\subsubsection{Ensembles vs.\ Learned Selection}
\label{sec:results-ensemble}
Beyond individual factors, we evaluate the utility of the mined libraries for downstream portfolio construction via two families of methods: simple factor combinations (equal-weight and IC-weighted) and learned selection models (Lasso and XGBoost trained on 2024 and evaluated on 2025). Across most baselines, learned models provide a clear uplift over simple ensembles, consistent with the presence of heterogeneous yet partially redundant signals in the candidate libraries. For FactorMiner, however, learned selection offers limited additional gains and can be slightly negative on some datasets (e.g., CSI500: EW and ICW ICIR = 1.29/1.31 vs.\ Lasso and XGBoost = 1.21/1.29), indicating that simple ensemble signals already capture most of the exploitable predictive power under the Train'24/Test'25 protocol (see \Cref{tab:main_results_kdd}).

\subsubsection{Additional Robustness Checks}
\label{sec:additional-robustness}
We further compare against end-to-end forecasting baselines that directly map the 10-minute market panel to predictive scores: PatchTST~\citep{nie2023patchtst}, Chronos-2~\citep{ansari2025chronos2}, and LightGBM~\citep{ke2017lightgbm}. These neural or tree-based predictors provide nontrivial signals, but they do not match the accuracy and stability of the interpretable formulaic library. As shown in \Cref{tab:e2e_baselines}, FactorMiner achieves stronger IC/ICIR than these end-to-end baselines across the evaluated markets. The gap is largest on A-shares, where LightGBM reaches 5.53\%/0.51 on CSI500 while FactorMiner reaches 8.25\%/0.77.

\begin{table}[htbp]
  \centering
  \caption{Comparison with end-to-end forecasting baselines on the 2025 test period. IC is in \%; higher is better.}
  \label{tab:e2e_baselines}
  \footnotesize
  \setlength{\tabcolsep}{3pt}
  \begin{tabular*}{\columnwidth}{@{\extracolsep{\fill}}l*{8}{c}@{}}
    \toprule
    \multirow{2}{*}{Market} & \multicolumn{2}{c}{PatchTST} & \multicolumn{2}{c}{Chronos-2} & \multicolumn{2}{c}{LightGBM} & \multicolumn{2}{c}{FactorMiner} \\
    \cmidrule(lr){2-3}\cmidrule(lr){4-5}\cmidrule(lr){6-7}\cmidrule(lr){8-9}
    & IC (\%) & ICIR & IC (\%) & ICIR & IC (\%) & ICIR & IC (\%) & ICIR \\
    \midrule
    CSI500  & 2.21 & 0.19 & 4.52 & 0.37 & 5.53 & 0.51 & \textbf{8.25} & \textbf{0.77} \\
    CSI1000 & 1.84 & 0.20 & 3.36 & 0.26 & 4.85 & 0.45 & \textbf{7.78} & \textbf{0.76} \\
    CSI300  & 1.80 & 0.15 & 3.01 & 0.19 & 4.66 & 0.34 & \textbf{7.46} & \textbf{0.38} \\
    Crypto  & 1.58 & 0.06 & 2.62 & 0.16 & 2.88 & 0.13 & \textbf{3.82} & \textbf{0.28} \\
    \bottomrule
  \end{tabular*}
\end{table}

The end-to-end signals are nevertheless complementary: their average absolute correlation with the released 110-factor library is below 0.04 and the maximum single-factor correlation is at most 0.23, while pairwise correlations among the three outputs stay modest (\Cref{app:e2e-corr}). Neural predictors may thus offer useful auxiliary information for future hybrid ensembles, while FactorMiner retains transparent, auditable formulas.

To test temporal persistence beyond the main 2025 evaluation window, we also evaluate the full 110-factor library on CSI500 over recent months through 2026-Q1. \Cref{tab:recent_robustness} shows that both the raw library and XGBoost-selected composite signal remain predictive in 2026-01 to 2026-03, supporting the view that the mined factors generalize beyond the original training year rather than merely fitting 2024-specific patterns. These results highlight a practical trade-off between factor timeliness and long-run stability: markets evolve, published anomalies may erode after becoming known and arbitraged~\citep{mclean2016academic}, and intraday microstructure signals can have shorter half-lives, making evidence under the current market regime especially important~\citep{aitsahalia2025highfreq}.

\begin{table}[htbp]
  \centering
  \caption{Recent-period robustness of the full factor library on CSI500.}
  \label{tab:recent_robustness}
  \footnotesize
  \setlength{\tabcolsep}{3pt}
  \begin{tabular*}{\columnwidth}{@{\extracolsep{\fill}}lcccc@{}}
    \toprule
    Period & Lib. IC (\%) & Lib. ICIR & XGB IC (\%) & XGB ICIR \\
    \midrule
    2025-10 & 5.52 & 0.49 & 13.42 & 1.16 \\
    2025-11 & 6.74 & 0.65 & 15.83 & 1.46 \\
    2025-12 & 6.83 & 0.68 & 16.16 & 1.55 \\
    2026-01 & 4.78 & 0.46 & 12.56 & 1.03 \\
    2026-02 & 6.23 & 0.64 & 15.23 & 1.38 \\
    2026-03 & 5.52 & 0.51 & 14.96 & 1.17 \\
    \bottomrule
  \end{tabular*}
\end{table}

The learned selector is also useful as a diagnostic tool for understanding which parts of the library carry nonlinear marginal value. \Cref{tab:xgb-importance} shows that XGBoost importance is not concentrated in a single VWAP-style anchor: the top factors span VWAP deviation, modernized Alpha101 formulas, momentum, regime switches, price-range interactions, divergence, higher moments, median-based transforms, and amount-efficiency signals. This broad support helps explain why tree-based selection improves several weaker libraries, while FactorMiner's own simple ensembles are already competitive because the admitted library is more diverse before learning.

\begin{table}[htbp]
  \centering
  \caption{XGBoost feature importance on the full FactorMiner library (top 20 factors).}
  \label{tab:xgb-importance}
  \scriptsize
  \setlength{\tabcolsep}{3pt}
  \begin{tabular*}{\columnwidth}{@{\extracolsep{\fill}}rllcl@{}}
    \toprule
    Rank & ID & Factor & Imp. & Category \\
    \midrule
     1 & 006 & VWAP Deviation & 6.04 & VWAP \\
     2 & 061 & Alpha101-12 Modern & 4.06 & Classic \\
     3 & 023 & Norm.-Momentum TsRank & 3.59 & Momentum \\
     4 & 068 & Skewness Regime PV-Div & 3.55 & Regime \\
     5 & 028 & Close-Low $\times$ Volume & 3.03 & Price range \\
     6 & 070 & Price-Pos. Vol. Interact. & 2.27 & Interaction \\
     7 & 057 & TsRank PV Divergence & 2.26 & Divergence \\
     8 & 029 & High-Close $\times$ Volume & 2.15 & Price range \\
     9 & 018 & Range-Position Vol. Regime & 1.62 & Range \\
    10 & 045 & Kurtosis-Regime Range & 1.57 & Higher moment \\
    11 & 048 & Vol-Price Rank Div. & 1.47 & Divergence \\
    12 & 104 & Median LogSwap & 1.46 & Median \\
    13 & 053 & Alpha101-1-V & 1.44 & Classic \\
    14 & 101 & Median Returns Switch & 1.41 & Median \\
    15 & 004 & High-Vol Rel. Weakness & 1.34 & Volume \\
    16 & 055 & Skew Open-Close Prod. & 1.24 & Distribution \\
    17 & 019 & VWAP-Gap Regime Range & 1.20 & VWAP \\
    18 & 092 & Amt Efficiency EMA & 1.16 & Efficiency \\
    19 & 073 & Amt Velocity Regime & 1.16 & Efficiency \\
    20 & 043 & Skewed Volume Momentum & 1.11 & Distribution \\
    \bottomrule
  \end{tabular*}
\end{table}

\subsection{Effect of Experience Memory}
\label{sec:ablation}
\begin{figure}[htbp]
  \centering
  \includegraphics[width=0.8\columnwidth]{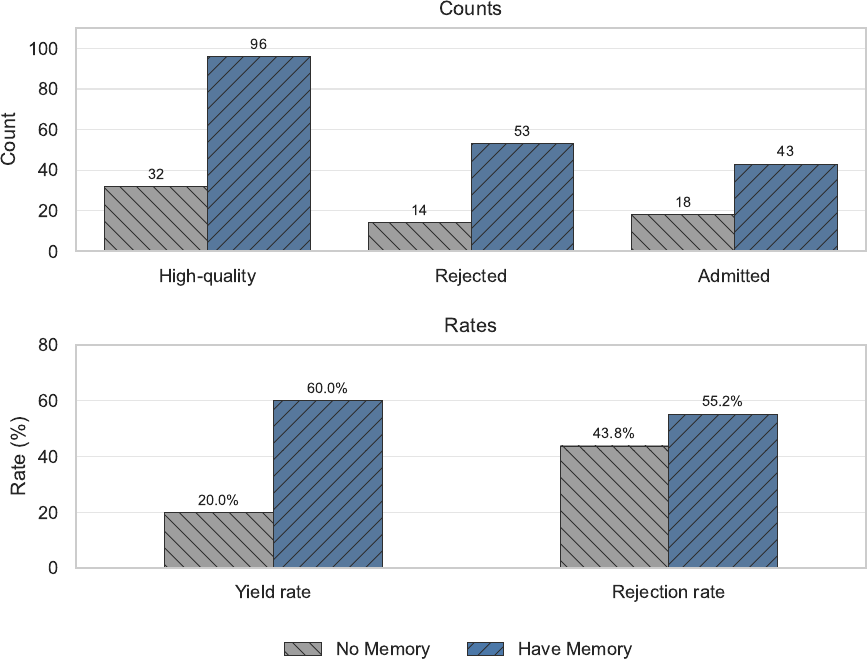}
  \vspace{-1ex}
  \caption{Ablation comparison between \textit{Have Memory} and No Memory. High-quality candidates who pass the IC threshold. The bar chart reports the counts and the corresponding yield and rejection rates.}
  \label{fig:ablation_memory_bar}
  \Description{}
  \vspace{-1ex}
\end{figure}

To isolate the contribution of the Experience Memory ($\mathcal{M}$), we conducted an ablation study comparing FactorMiner against a "No Memory" variant where operators $F, E, R$ were disabled, reducing the system to standard LLM-based evolution (see \Cref{fig:ablation_memory_bar} for a summary).
Since this experiment is designed as a controlled mechanism ablation, we adopt relatively relaxed screening thresholds to ensure sufficient samples for comparison (IC threshold $|\text{IC}|>0.02$; redundancy threshold $\theta=0.85$ for this ablation setting). \Cref{fig:ablation_memory_bar} highlights the impact of memory guidance:
{\setlength{\leftmargini}{1.2em}%
\begin{itemize}
    \item \textbf{Precise Navigation (High Yield)}: The \textit{Have Memory} variant generates 96 high-quality candidates (60.0\% yield), whereas the No Memory baseline produces only 32 (20.0\% yield). This suggests that the Retrieval operator effectively maps the search space, allowing the agent to guide exploration toward regions with higher expected yield rather than exploring randomly.
    \item \textbf{Aggressive Filtration (High Diversity)}: Despite generating $3.0\times$ more valid signals, the \textit{Have Memory} variant actively rejects a higher proportion of them for redundancy (55.2\% vs 43.8\%). This confirms that the Evolution operator ($E$) functions as a strategic filter, prioritizing unique signal discovery over mere quantity.
\end{itemize}
}

Beyond aggregate counts, the retrieved memories expose reusable search priors rather than opaque prompts. \Cref{tab:recommended-directions} summarizes representative directions that repeatedly led to admitted, low redundancy factors. These priors make the search less trial-and-error: the agent is steered toward mechanism families such as higher moment regimes, amount efficiency, and trend regression adaptivity, while later memory updates can suppress directions that collapse back into crowded VWAP deviation clusters.

\begin{table}[htbp]
  \centering
  \caption{Recommended mining directions distilled from experience memory.}
  \label{tab:recommended-directions}
  \scriptsize
  \setlength{\tabcolsep}{2pt}
  \begin{tabular}{@{}p{0.31\columnwidth}p{0.52\columnwidth}c@{}}
    \toprule
    Pattern & Search prior & Yield \\
    \midrule
    Higher moments & Use Skew/Kurt as regime conditions for reversal logic. & High \\
    PV corr. interaction & Combine price-volume correlation with amount or trend operators. & High \\
    Robust efficiency & Smooth return/amount efficiency using median or robust statistics. & High \\
    Smoothed efficiency rank & Apply EMA/WMA before cross-sectional ranking. & High \\
    Trend regression & Switch between Slope and Resi using Rsquare reliability. & High \\
    Extreme regimes & Use logical Or over price/volume extremes as state triggers. & High \\
    Kurtosis regime & Adapt reversal windows under fat-tail environments. & High \\
    Amt-efficiency interaction & Cross amount-efficiency rank with Skew/Kurt features. & Med. \\
    \bottomrule
  \end{tabular}
\end{table}

\subsection{Mining Efficiency Analysis}
\label{sec:efficiency}

To demonstrate the practical advantage of the factor mining skill, we conduct rigorous benchmarks comparing three execution backends supported by our framework: (1) standard \textbf{Python (Pandas)}, (2) \textbf{C-compiled (Bottleneck)} for efficient CPU execution, and (3) \textbf{GPU-accelerated} for massive parallelism. This multi-backend support allows the agent to remain lightweight while achieving industrial-grade efficiency through on-demand acceleration.

\begin{figure}[htbp]
  \centering
  \includegraphics[width=0.8\columnwidth]{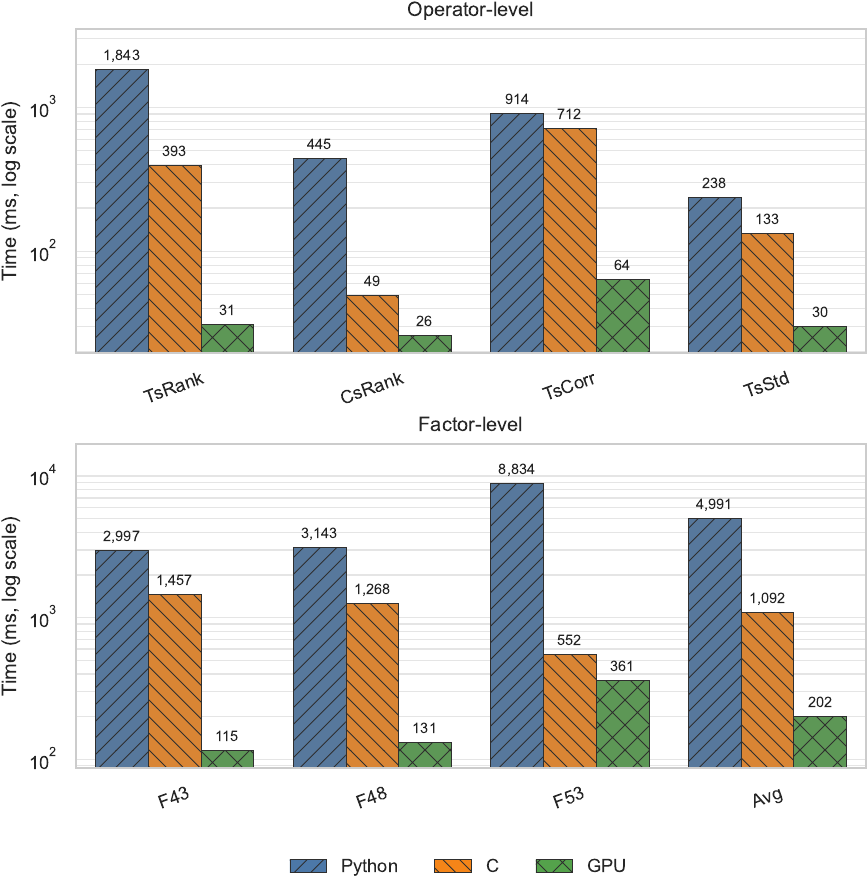}
  \vspace{-0.5ex}
  \caption{Operator and factor level computation time (ms, log scale) under Python, optimized CPU (C), and GPU backends; lower is better. The GPU backend yields consistent order-of-magnitude gains.}
  \label{fig:efficiency-bench-bar}
  \Description{}
  \vspace{-1ex}
\end{figure}

\textbf{Results.}
\Cref{fig:efficiency-bench-bar} reports both operator-level and factor-level benchmarks on the CSI500 dataset (12,610 $\times$ 500), measuring \textit{pure computation time} excluding I/O. At the operator level, the GPU backend achieves \textbf{8--59$\times$} speedups over Pandas and \textbf{2--13$\times$} over the optimized C backend. The C backend is valuable for lightweight CPU-only runs, while the GPU backend is decisive for repeated large-batch evaluation.

For end-to-end factor evaluation (\Cref{fig:efficiency-bench-bar}, factor level), rank-intensive factors (F43, F48, F53) show substantial \textbf{23--27$\times$} speedups, and even the highly-optimized C implementation is \textbf{5.4$\times$} slower than the GPU backend on average (1,092ms vs. 202ms). By offloading these intensive primitives to GPU/C while parallelizing batch evaluation, FactorMiner keeps large-scale iterative discovery feasible on commodity hardware: evaluating 1,000 candidate factors takes $\sim$6 minutes versus $\sim$70 minutes with Pandas, keeping the iterative "generate-evaluate-refine" loop practical.

\section{Discussion}

High-frequency markets can be treated as a data-rich complex adaptive system, and each formulaic factor can be viewed as an explicit, falsifiable hypothesis about market microstructure. Beyond predictive performance, FactorMiner yields 110 interpretable high-frequency alpha factors and a standardized evaluation protocol for reproducible, hypothesis-driven analysis of market microstructure.

\paragraph{Experience memory as continual learning.}
Our results confirm that memory-guided, skill-based agents can scale neural-symbolic program synthesis while retaining interpretability.
The experience memory acts as institutional knowledge accumulated across sessions, enabling meta-learning: the system learns not just individual factors but how to search more effectively.
Key insights include \textbf{successful patterns} (higher-moment regimes via Skew/Kurt, trend-regression adaptivity via Rsquare/Slope/Resi, amount-efficiency interactions) that yield high-IC low-correlation factors; \textbf{failure patterns} (VWAP-deviation variants, standardized returns, simple Delta reversals) that correlate with existing factors; and the \textbf{Correlation Red Sea}, where new orthogonal factors get harder to find as the library grows without memory guidance.

\section{Conclusion}
FactorMiner provides a lightweight self-evolving agent framework for interpretable high-frequency alpha discovery by combining a modular mining skill with experience memory. The resulting 110-factor library and standardized protocol form a reproducible artifact that supports hypothesis testing, mechanistic inspection, and cross-market transfer studies of market microstructure. Future work will incorporate transaction-cost-aware backtesting, extend to broader assets and frequencies, and develop online memory updates for non-stationary markets.

\section{Limitations and Ethical Considerations}
Although we include comparisons with end-to-end models, these baselines are not exhaustively tuned and remain less interpretable than formulaic factors. Our pipeline also targets predictive IC, ICIR, redundancy, and portfolio diagnostics, with transaction-cost-aware deployment left for future work. The discovered factors could be misused in speculative or manipulative strategies; deployment should follow compliance and risk-control requirements.

\begin{acks}
This work was supported in part by the Shenzhen Ubiquitous Data Enabling Key Lab under Grant ZDSYS20220527171406015, and in part by the Tsinghua Shenzhen International Graduate School-Shenzhen Pengrui Endowed Professorship Scheme of Shenzhen Pengrui Foundation.
\end{acks}
\bibliographystyle{ACM-Reference-Format}
\bibliography{factorminer}

\appendix

\section{Operator Library and Admission Criteria}
\label{app:factor-details}
\label{app:admission}

Each symbolic factor $\alpha \in \mathcal{P}$ is an expression tree built from raw feature references (\texttt{\$open}, \texttt{\$high}, \texttt{\$low}, \texttt{\$close}, \texttt{\$volume}, \texttt{\$amt}, \texttt{\$vwap}, \texttt{\$returns}) and a curated library of 60+ typed operators, organized into the categories shown in \Cref{tab:operators}; \Cref{tab:operators} lists representative operators for readability, while the full FactorMiner registry contains 60+ typed operators.

\begin{table}[hb]
  \centering
  \caption{Operator categories in the factor mining skill (representative operators; 60+ in total).}
  \label{tab:operators}
  \scriptsize
  \setlength{\tabcolsep}{3pt}
  \begin{tabular}{@{}l p{0.82\columnwidth}@{}}
    \toprule
    Category & Representative operators \\
    \midrule
    Arithmetic & Add, Sub, Mul, Div, Neg, Abs, Log, SignedPower, Power, Inv, Sqrt, Square, Exp, Tanh \\
    Statistical & Mean, Std, Var, Skew, Kurt, Med, Sum, Product \\
    Time-series & Delay, Delta, TsRank, TsMax, TsMin, TsArgMax, TsArgMin, TsDecay \\
    Cross-sectional & CsRank, Scale \\
    Smoothing & SMA, EMA, WMA \\
    Regression & Slope, Rsquare, Resi \\
    Logical & IfElse, Greater, Less, GreaterEqual, LessEqual, And, Or, Eq, Ne \\
    \bottomrule
  \end{tabular}
\end{table}

\begingroup
\setlength{\abovedisplayskip}{3pt}\setlength{\belowdisplayskip}{3pt}
\setlength{\abovedisplayshortskip}{2pt}\setlength{\belowdisplayshortskip}{2pt}
\noindent\textbf{Admission and replacement.}
A symbolic factor $\alpha$ is admitted to the library $\mathcal{L}$ if it satisfies
\begin{equation}
  \text{IC}(\alpha) \geq \tau_{\text{IC}} \quad \land \quad \max_{g \in \mathcal{L}} |\rho(\alpha, g)| < \theta,
\end{equation}
where $\tau_{\text{IC}}$ and $\theta$ are the IC and correlation thresholds (defaults $\tau_{\text{IC}}=0.04$ and $\theta=0.5$ for A-share mining unless otherwise specified). We further introduce a factor replacement mechanism for high-quality candidates that would otherwise be rejected due to correlation: if a new factor $\alpha$ satisfies
\begin{equation}
  \text{IC}(\alpha) \geq 0.10 \;\;\land\;\; \text{IC}(\alpha) \geq 1.3\,\text{IC}(g) \;\;\land\;\; |\{g \in \mathcal{L} : \rho(\alpha, g) > \theta\}| = 1,
\end{equation}
then $\alpha$ replaces the single correlated factor $g$, allowing the library to evolve toward higher quality while preserving diversity.
\par\endgroup

\noindent\textbf{Reproducibility.}
Symbolic proposals are generated by Gemini 3.0 Flash without any weight updates. Factor evaluation uses GPU/C-accelerated operators (NumPy/CuPy) and a 40-worker multiprocessing pool; representative operator runtimes are reported in \Cref{fig:efficiency-bench-bar}. Alpha101 baselines, originally daily-frequency, are adapted to 10-minute bars by generating up to 10 window-parameter variants per formula and keeping the best configuration.

\section{End-to-End Signal Correlations}
\label{app:e2e-corr}

We test whether the end-to-end baselines (PatchTST, Chronos-2, LightGBM) are redundant with the released 110-factor library. \Cref{tab:app-e2e-library-corr} shows the overlap is small: average absolute correlations are below 0.04 and the maximum single-factor correlation is at most 0.23. \Cref{tab:app-e2e-pairwise-corr} shows that pairwise correlations among the three signals are also modest. These models do not dominate the formulaic library in predictive performance, but their outputs may carry complementary information for future hybrid systems.

\begin{table}[t]
  \centering
  \caption{Correlation between end-to-end model signals and the 110-factor library.}
  \label{tab:app-e2e-library-corr}
  \footnotesize
  \setlength{\tabcolsep}{1.5pt}
  \begin{tabular*}{\columnwidth}{@{\extracolsep{\fill}}lccp{0.31\columnwidth}cc@{}}
    \toprule
    Model & Avg $|\rho|$ & Max $|\rho|$ & Nearest factor & 2nd & 3rd \\
    \midrule
    PatchTST & 0.0365 & 0.18 & 002 EMA Deviation & 108 & 006 \\
    Chronos-2 & 0.0399 & 0.23 & 062 Amount-stability adj. returns & 002 & 081 \\
    LightGBM & 0.0146 & 0.08 & 062 Amount-stability adj. returns & 006 & 084 \\
    \bottomrule
  \end{tabular*}
\end{table}

\begin{table}[t]
  \centering
  \caption{Pairwise correlations among end-to-end signals.}
  \label{tab:app-e2e-pairwise-corr}
  \footnotesize
  \setlength{\tabcolsep}{5pt}
  \begin{tabular*}{\columnwidth}{@{\extracolsep{\fill}}lccc@{}}
    \toprule
    Model & Chronos-2 & LightGBM & PatchTST \\
    \midrule
    Chronos-2 & 1.000 & 0.043 & 0.171 \\
    LightGBM & 0.043 & 1.000 & 0.224 \\
    PatchTST & 0.171 & 0.224 & 1.000 \\
    \bottomrule
  \end{tabular*}
\end{table}

\section{Experience Memory: Forbidden Directions}
\label{app:forbidden}

Beyond successful templates, the memory also distills \emph{negative knowledge}: structural directions that repeatedly produce candidates highly correlated with the existing library. \Cref{tab:forbidden} summarizes the main forbidden directions and their typical correlation with admitted factors; recording them prevents the agent from wasting budget rediscovering known signals.

\begin{table}[h]
  \centering
  \caption{Forbidden mining directions (high correlation risk).}
  \label{tab:forbidden}
  \footnotesize
  \setlength{\tabcolsep}{3pt}
  \begin{tabular}{@{}p{0.44\columnwidth}p{0.34\columnwidth}c@{}}
    \toprule
    Direction & Correlated factors & $|\rho|$ \\
    \midrule
    Standardized Returns/Amount & 006, 008, 009 & $>$0.6 \\
    VWAP Deviation variants & 006, 009, 012, 013, 016 & $>$0.5 \\
    Mean Reversion / Price Deviation & 001, 002 & $>$0.5 \\
    Simple Delta Reversal & 023, 024 & $>$0.5 \\
    Close-Position Location & 028, 044 & 0.87+ \\
    Volatility + Price Position Branch & 046, 064 & $>$0.6 \\
    High R$^2$ Trend Following & 081, 083 & $>$0.6 \\
    Rsquare Weighted Momentum & 002, 023 & $>$0.7 \\
    WMA/EMA Smoothed Efficiency & 092 & $>$0.9 \\
    \bottomrule
  \end{tabular}
\end{table}

The VWAP cluster (factor 006 and variants) is the densest correlation region: close-to-VWAP signals normalized by volatility or volume almost always correlate $>$0.5 with it. As the library grew past 70 factors, even high-IC candidates increasingly collided with existing ones, indicating the ``easy'' signal space was largely exhausted.

\section{Full Factor Library (110 factors)}
\label{app:factor-list}

\begingroup
\scriptsize
\linespread{0.95}\selectfont
\raggedright
\setlength{\parindent}{0pt}
\setlength{\parskip}{0.1pt}
{\noindent\rule{\columnwidth}{0.6pt}\par\nobreak\vskip1pt
\noindent\makebox[2.8em][l]{\textbf{ID}}\textbf{Formula}\par\nobreak\vskip1pt
\noindent\rule{\columnwidth}{0.4pt}\par}
\vskip2pt
\newcommand{\frow}[2]{\par\hangindent=2.9em\hangafter=1\noindent\makebox[2.6em][l]{\texttt{#1}}#2}
\frow{001}{\texttt{Neg(\allowbreak CsRank(\allowbreak Div(\allowbreak Sub(\allowbreak \$close, Min(\allowbreak \$close, 48)\allowbreak )\allowbreak , Add(\allowbreak Sub(\allowbreak Max(\allowbreak \$close, 48)\allowbreak , Min(\allowbreak \$close, 48)\allowbreak )\allowbreak , 1e-8)\allowbreak )\allowbreak )\allowbreak )\allowbreak }}
\frow{002}{\texttt{Neg(\allowbreak Div(\allowbreak Sub(\allowbreak \$close, EMA(\allowbreak \$close, 10)\allowbreak )\allowbreak , EMA(\allowbreak \$close, 10)\allowbreak )\allowbreak )\allowbreak }}
\frow{003}{\texttt{Sub(\allowbreak CsRank(\allowbreak Delta(\allowbreak \$volume, 1)\allowbreak )\allowbreak , CsRank(\allowbreak Div(\allowbreak Sub(\allowbreak \$close, \$vwap)\allowbreak , \$vwap)\allowbreak )\allowbreak )\allowbreak }}
\frow{004}{\texttt{Mul(\allowbreak CsRank(\allowbreak Div(\allowbreak \$volume, Mean(\allowbreak \$volume, 24)\allowbreak )\allowbreak )\allowbreak , CsRank(\allowbreak Neg(\allowbreak \$returns)\allowbreak )\allowbreak )\allowbreak }}
\frow{005}{\texttt{Neg(\allowbreak Mul(\allowbreak TsRank(\allowbreak Div(\allowbreak Sub(\allowbreak \$close, TsMin(\allowbreak \$close, 12)\allowbreak )\allowbreak , Add(\allowbreak Sub(\allowbreak TsMax(\allowbreak \$close, 12)\allowbreak , TsMin(\allowbreak \$close, 12)\allowbreak )\allowbreak , 1e-6)\allowbreak )\allowbreak , 12)\allowbreak , CsRank(\allowbreak Abs(\allowbreak Cov(\allowbreak \$returns, \$volume, 12)\allowbreak )\allowbreak )\allowbreak )\allowbreak )\allowbreak }}
\frow{006}{\texttt{Neg(\allowbreak Div(\allowbreak Sub(\allowbreak \$close, \$vwap)\allowbreak , \$vwap)\allowbreak )\allowbreak }}
\frow{007}{\texttt{Neg(\allowbreak Sub(\allowbreak CsRank(\allowbreak Div(\allowbreak Sub(\allowbreak \$close, \$low)\allowbreak , Add(\allowbreak Sub(\allowbreak \$high, \$low)\allowbreak , 0.0001)\allowbreak )\allowbreak )\allowbreak , CsRank(\allowbreak Delta(\allowbreak EMA(\allowbreak \$volume, 5)\allowbreak , 5)\allowbreak )\allowbreak )\allowbreak )\allowbreak }}
\frow{008}{\texttt{Neg(\allowbreak Mul(\allowbreak CsRank(\allowbreak Div(\allowbreak \$returns, Add(\allowbreak Std(\allowbreak \$returns, 12)\allowbreak , 0.0001)\allowbreak )\allowbreak )\allowbreak , CsRank(\allowbreak Div(\allowbreak \$volume, Mean(\allowbreak \$volume, 12)\allowbreak )\allowbreak )\allowbreak )\allowbreak )\allowbreak }}
\frow{009}{\texttt{Neg(\allowbreak Mul(\allowbreak CsRank(\allowbreak \$returns)\allowbreak , CsRank(\allowbreak Std(\allowbreak \$returns, 12)\allowbreak )\allowbreak )\allowbreak )\allowbreak }}
\frow{010}{\texttt{Neg(\allowbreak CsRank(\allowbreak Sub(\allowbreak CsRank(\allowbreak Div(\allowbreak \$close, Delay(\allowbreak \$close, 5)\allowbreak )\allowbreak )\allowbreak , CsRank(\allowbreak Div(\allowbreak \$volume, Delay(\allowbreak \$volume, 5)\allowbreak )\allowbreak )\allowbreak )\allowbreak )\allowbreak )\allowbreak }}
\frow{011}{\texttt{Neg(\allowbreak Sub(\allowbreak CsRank(\allowbreak Delta(\allowbreak \$close, 6)\allowbreak )\allowbreak , CsRank(\allowbreak Delta(\allowbreak \$vwap, 6)\allowbreak )\allowbreak )\allowbreak )\allowbreak }}
\frow{012}{\texttt{Mul(\allowbreak CsRank(\allowbreak Neg(\allowbreak Div(\allowbreak Sub(\allowbreak \$close, \$vwap)\allowbreak , \$vwap)\allowbreak )\allowbreak )\allowbreak , CsRank(\allowbreak Div(\allowbreak \$volume, EMA(\allowbreak \$volume, 12)\allowbreak )\allowbreak )\allowbreak )\allowbreak }}
\frow{013}{\texttt{Mul(\allowbreak CsRank(\allowbreak Neg(\allowbreak Div(\allowbreak Sub(\allowbreak \$close, \$vwap)\allowbreak , \$vwap)\allowbreak )\allowbreak )\allowbreak , Sub(\allowbreak 1, CsRank(\allowbreak \$volume)\allowbreak )\allowbreak )\allowbreak }}
\frow{014}{\texttt{Mul(\allowbreak Add(\allowbreak Mul(\allowbreak CsRank(\allowbreak Neg(\allowbreak Div(\allowbreak Sub(\allowbreak \$close, \$vwap)\allowbreak , \$vwap)\allowbreak )\allowbreak )\allowbreak , 0.6)\allowbreak , Mul(\allowbreak CsRank(\allowbreak Delta(\allowbreak Neg(\allowbreak Div(\allowbreak Sub(\allowbreak \$close, \$vwap)\allowbreak , \$vwap)\allowbreak )\allowbreak , 3)\allowbreak )\allowbreak , 0.4)\allowbreak )\allowbreak , CsRank(\allowbreak Delta(\allowbreak Div(\allowbreak Mean(\allowbreak Div(\allowbreak Abs(\allowbreak \$returns)\allowbreak , Add(\allowbreak \$volume, 1)\allowbreak )\allowbreak , 24)\allowbreak , Add(\allowbreak Div(\allowbreak Abs(\allowbreak \$returns)\allowbreak , Add(\allowbreak \$volume, 1)\allowbreak )\allowbreak , 1e-6)\allowbreak )\allowbreak , 3)\allowbreak )\allowbreak )\allowbreak }}
\frow{015}{\texttt{Mul(\allowbreak CsRank(\allowbreak Neg(\allowbreak Delta(\allowbreak \$returns, 3)\allowbreak )\allowbreak )\allowbreak , Mul(\allowbreak Sub(\allowbreak 1, CsRank(\allowbreak Std(\allowbreak \$returns, 12)\allowbreak )\allowbreak )\allowbreak , CsRank(\allowbreak Std(\allowbreak \$returns, 12)\allowbreak )\allowbreak )\allowbreak )\allowbreak }}
\frow{016}{\texttt{Neg(\allowbreak Mul(\allowbreak CsRank(\allowbreak Delta(\allowbreak Div(\allowbreak Sub(\allowbreak \$close, \$vwap)\allowbreak , \$vwap)\allowbreak , 3)\allowbreak )\allowbreak , CsRank(\allowbreak Div(\allowbreak \$volume, EMA(\allowbreak \$volume, 12)\allowbreak )\allowbreak )\allowbreak )\allowbreak )\allowbreak }}
\frow{017}{\texttt{Neg(\allowbreak Mul(\allowbreak TsRank(\allowbreak Delta(\allowbreak Div(\allowbreak Sub(\allowbreak \$close, \$vwap)\allowbreak , \$vwap)\allowbreak , 3)\allowbreak , 12)\allowbreak , TsRank(\allowbreak Std(\allowbreak \$returns, 12)\allowbreak , 12)\allowbreak )\allowbreak )\allowbreak }}
\frow{018}{\texttt{Neg(\allowbreak Mul(\allowbreak TsRank(\allowbreak Div(\allowbreak Sub(\allowbreak \$close, TsMin(\allowbreak \$close, 12)\allowbreak )\allowbreak , Add(\allowbreak Sub(\allowbreak TsMax(\allowbreak \$close, 12)\allowbreak , TsMin(\allowbreak \$close, 12)\allowbreak )\allowbreak , 1e-6)\allowbreak )\allowbreak , 12)\allowbreak , TsRank(\allowbreak Div(\allowbreak \$volume, EMA(\allowbreak \$volume, 12)\allowbreak )\allowbreak , 12)\allowbreak )\allowbreak )\allowbreak }}
\frow{019}{\texttt{IfElse(\allowbreak Greater(\allowbreak Abs(\allowbreak Div(\allowbreak Sub(\allowbreak \$close, \$vwap)\allowbreak , \$vwap)\allowbreak )\allowbreak , 0.01)\allowbreak , Neg(\allowbreak CsRank(\allowbreak Sub(\allowbreak \$close, TsMax(\allowbreak \$close, 12)\allowbreak )\allowbreak )\allowbreak )\allowbreak , Neg(\allowbreak CsRank(\allowbreak Sub(\allowbreak \$close, TsMin(\allowbreak \$close, 12)\allowbreak )\allowbreak )\allowbreak )\allowbreak )\allowbreak }}
\frow{020}{\texttt{TsRank(\allowbreak Sub(\allowbreak CsRank(\allowbreak Delta(\allowbreak \$volume, 3)\allowbreak )\allowbreak , CsRank(\allowbreak Delta(\allowbreak \$close, 3)\allowbreak )\allowbreak )\allowbreak , 18)\allowbreak }}
\frow{021}{\texttt{Neg(\allowbreak Mul(\allowbreak TsRank(\allowbreak Delta(\allowbreak Div(\allowbreak \$volume, EMA(\allowbreak \$volume, 12)\allowbreak )\allowbreak , 3)\allowbreak , 12)\allowbreak , TsRank(\allowbreak Delta(\allowbreak Div(\allowbreak Sub(\allowbreak \$close, \$vwap)\allowbreak , \$vwap)\allowbreak , 3)\allowbreak , 12)\allowbreak )\allowbreak )\allowbreak }}
\frow{022}{\texttt{Neg(\allowbreak CsRank(\allowbreak Div(\allowbreak Sub(\allowbreak Min2(\allowbreak \$open, \$close)\allowbreak , \$low)\allowbreak , Add(\allowbreak Sub(\allowbreak \$high, \$low)\allowbreak , 0.0001)\allowbreak )\allowbreak )\allowbreak )\allowbreak }}
\frow{023}{\texttt{Neg(\allowbreak TsRank(\allowbreak Div(\allowbreak Delta(\allowbreak \$close, 3)\allowbreak , Mean(\allowbreak Abs(\allowbreak Delta(\allowbreak \$close, 3)\allowbreak )\allowbreak , 12)\allowbreak )\allowbreak , 12)\allowbreak )\allowbreak }}
\frow{024}{\texttt{Sub(\allowbreak TsRank(\allowbreak Delta(\allowbreak \$open, 6)\allowbreak , 12)\allowbreak , TsRank(\allowbreak Delta(\allowbreak \$close, 6)\allowbreak , 12)\allowbreak )\allowbreak }}
\frow{025}{\texttt{Sub(\allowbreak TsRank(\allowbreak Delta(\allowbreak \$open, 9)\allowbreak , 18)\allowbreak , TsRank(\allowbreak Delta(\allowbreak \$close, 9)\allowbreak , 18)\allowbreak )\allowbreak }}
\frow{026}{\texttt{Neg(\allowbreak Mul(\allowbreak TsRank(\allowbreak Cov(\allowbreak TsRank(\allowbreak \$close, 18)\allowbreak , TsRank(\allowbreak \$volume, 18)\allowbreak , 10)\allowbreak , 18)\allowbreak , TsRank(\allowbreak Div(\allowbreak Sub(\allowbreak \$close, \$low)\allowbreak , Sub(\allowbreak \$high, \$low)\allowbreak )\allowbreak , 18)\allowbreak )\allowbreak )\allowbreak }}
\frow{027}{\texttt{Neg(\allowbreak Mul(\allowbreak TsRank(\allowbreak Cov(\allowbreak TsRank(\allowbreak \$open, 12)\allowbreak , TsRank(\allowbreak \$volume, 12)\allowbreak , 5)\allowbreak , 12)\allowbreak , TsRank(\allowbreak Div(\allowbreak Sub(\allowbreak \$close, \$low)\allowbreak , Sub(\allowbreak \$high, \$low)\allowbreak )\allowbreak , 12)\allowbreak )\allowbreak )\allowbreak }}
\frow{028}{\texttt{Neg(\allowbreak Mul(\allowbreak Sub(\allowbreak \$close, \$low)\allowbreak , Div(\allowbreak \$volume, Mean(\allowbreak \$volume, 12)\allowbreak )\allowbreak )\allowbreak )\allowbreak }}
\frow{029}{\texttt{Mul(\allowbreak Sub(\allowbreak \$high, \$close)\allowbreak , Div(\allowbreak \$volume, Mean(\allowbreak \$volume, 12)\allowbreak )\allowbreak )\allowbreak }}
\frow{030}{\texttt{Mul(\allowbreak Delta(\allowbreak Sub(\allowbreak \$high, \$close)\allowbreak , 3)\allowbreak , Div(\allowbreak \$volume, Mean(\allowbreak \$volume, 12)\allowbreak )\allowbreak )\allowbreak }}
\frow{031}{\texttt{IfElse(\allowbreak Greater(\allowbreak Delta(\allowbreak \$close, 1)\allowbreak , 0)\allowbreak , Neg(\allowbreak TsRank(\allowbreak Delta(\allowbreak \$close, 3)\allowbreak , 12)\allowbreak )\allowbreak , TsRank(\allowbreak Delta(\allowbreak \$close, 3)\allowbreak , 12)\allowbreak )\allowbreak }}
\frow{032}{\texttt{Neg(\allowbreak TsRank(\allowbreak Sub(\allowbreak CsRank(\allowbreak Delta(\allowbreak \$returns, 2)\allowbreak )\allowbreak , CsRank(\allowbreak Delta(\allowbreak \$volume, 2)\allowbreak )\allowbreak )\allowbreak , 12)\allowbreak )\allowbreak }}
\frow{033}{\texttt{Neg(\allowbreak Mul(\allowbreak TsRank(\allowbreak Corr(\allowbreak TsRank(\allowbreak Div(\allowbreak \$volume, Mean(\allowbreak \$volume, 24)\allowbreak )\allowbreak , 12)\allowbreak , TsRank(\allowbreak Sub(\allowbreak \$high, \$low)\allowbreak , 12)\allowbreak , 12)\allowbreak , 24)\allowbreak , CsRank(\allowbreak Delta(\allowbreak \$close, 2)\allowbreak )\allowbreak )\allowbreak )\allowbreak }}
\frow{034}{\texttt{IfElse(\allowbreak Greater(\allowbreak Delta(\allowbreak \$close, 2)\allowbreak , 0)\allowbreak , Neg(\allowbreak TsRank(\allowbreak Delta(\allowbreak \$close, 5)\allowbreak , 15)\allowbreak )\allowbreak , TsRank(\allowbreak Delta(\allowbreak \$close, 5)\allowbreak , 15)\allowbreak )\allowbreak }}
\frow{035}{\texttt{Neg(\allowbreak Mul(\allowbreak TsRank(\allowbreak Delta(\allowbreak Sub(\allowbreak \$high, \$low)\allowbreak , 2)\allowbreak , 12)\allowbreak , CsRank(\allowbreak Delta(\allowbreak \$close, 2)\allowbreak )\allowbreak )\allowbreak )\allowbreak }}
\frow{036}{\texttt{IfElse(\allowbreak Greater(\allowbreak Std(\allowbreak \$returns, 12)\allowbreak , Mean(\allowbreak Std(\allowbreak \$returns, 12)\allowbreak , 48)\allowbreak )\allowbreak , Neg(\allowbreak CsRank(\allowbreak Div(\allowbreak Sub(\allowbreak Min2(\allowbreak \$open, \$close)\allowbreak , \$low)\allowbreak , Add(\allowbreak Sub(\allowbreak \$high, \$low)\allowbreak , 0.0001)\allowbreak )\allowbreak )\allowbreak )\allowbreak , Neg(\allowbreak CsRank(\allowbreak \$returns)\allowbreak )\allowbreak )\allowbreak }}
\frow{037}{\texttt{Neg(\allowbreak TsRank(\allowbreak Mul(\allowbreak CsRank(\allowbreak Delta(\allowbreak \$close, 3)\allowbreak )\allowbreak , CsRank(\allowbreak Delta(\allowbreak \$volume, 3)\allowbreak )\allowbreak )\allowbreak , 18)\allowbreak )\allowbreak }}
\frow{038}{\texttt{IfElse(\allowbreak Greater(\allowbreak Std(\allowbreak \$returns, 12)\allowbreak , Mean(\allowbreak Std(\allowbreak \$returns, 12)\allowbreak , 48)\allowbreak )\allowbreak , Neg(\allowbreak TsRank(\allowbreak Sub(\allowbreak CsRank(\allowbreak Delta(\allowbreak \$volume, 3)\allowbreak )\allowbreak , CsRank(\allowbreak Delta(\allowbreak \$close, 3)\allowbreak )\allowbreak )\allowbreak , 18)\allowbreak )\allowbreak , Neg(\allowbreak TsRank(\allowbreak Div(\allowbreak Sub(\allowbreak \$close, \$vwap)\allowbreak , Add(\allowbreak Std(\allowbreak \$returns, 24)\allowbreak , 0.0001)\allowbreak )\allowbreak , 12)\allowbreak )\allowbreak )\allowbreak }}
\frow{039}{\texttt{Neg(\allowbreak Mul(\allowbreak CsRank(\allowbreak Delta(\allowbreak \$close, 5)\allowbreak )\allowbreak , CsRank(\allowbreak Kurt(\allowbreak \$returns, 24)\allowbreak )\allowbreak )\allowbreak )\allowbreak }}
\frow{040}{\texttt{Neg(\allowbreak Mul(\allowbreak CsRank(\allowbreak Div(\allowbreak Sub(\allowbreak \$close, \$low)\allowbreak , Add(\allowbreak Sub(\allowbreak \$high, \$low)\allowbreak , 0.0001)\allowbreak )\allowbreak )\allowbreak , CsRank(\allowbreak Skew(\allowbreak \$returns, 24)\allowbreak )\allowbreak )\allowbreak )\allowbreak }}
\frow{041}{\texttt{Mul(\allowbreak CsRank(\allowbreak Div(\allowbreak Sub(\allowbreak \$high, \$close)\allowbreak , Add(\allowbreak Sub(\allowbreak \$high, \$low)\allowbreak , 1e-8)\allowbreak )\allowbreak )\allowbreak , CsRank(\allowbreak Neg(\allowbreak Skew(\allowbreak \$returns, 24)\allowbreak )\allowbreak )\allowbreak )\allowbreak }}
\frow{042}{\texttt{IfElse(\allowbreak Greater(\allowbreak Abs(\allowbreak Skew(\allowbreak \$returns, 24)\allowbreak )\allowbreak , 1.0)\allowbreak , Neg(\allowbreak CsRank(\allowbreak \$returns)\allowbreak )\allowbreak , Neg(\allowbreak CsRank(\allowbreak Skew(\allowbreak \$returns, 24)\allowbreak )\allowbreak )\allowbreak )\allowbreak }}
\frow{043}{\texttt{Neg(\allowbreak TsRank(\allowbreak Mul(\allowbreak CsRank(\allowbreak Skew(\allowbreak \$volume, 24)\allowbreak )\allowbreak , CsRank(\allowbreak Delta(\allowbreak \$close, 3)\allowbreak )\allowbreak )\allowbreak , 12)\allowbreak )\allowbreak }}
\frow{044}{\texttt{Neg(\allowbreak Mul(\allowbreak CsRank(\allowbreak Kurt(\allowbreak \$volume, 24)\allowbreak )\allowbreak , CsRank(\allowbreak Div(\allowbreak Sub(\allowbreak \$close, \$low)\allowbreak , Add(\allowbreak Sub(\allowbreak \$high, \$low)\allowbreak , 0.0001)\allowbreak )\allowbreak )\allowbreak )\allowbreak )\allowbreak }}
\frow{045}{\texttt{IfElse(\allowbreak Greater(\allowbreak Kurt(\allowbreak \$returns, 24)\allowbreak , 3.0)\allowbreak , Neg(\allowbreak CsRank(\allowbreak Div(\allowbreak Sub(\allowbreak \$high, \$close)\allowbreak , Add(\allowbreak Sub(\allowbreak \$high, \$low)\allowbreak , 0.0001)\allowbreak )\allowbreak )\allowbreak )\allowbreak , Neg(\allowbreak CsRank(\allowbreak Div(\allowbreak Sub(\allowbreak \$close, \$low)\allowbreak , Add(\allowbreak Sub(\allowbreak \$high, \$low)\allowbreak , 0.0001)\allowbreak )\allowbreak )\allowbreak )\allowbreak )\allowbreak }}
\frow{046}{\texttt{IfElse(\allowbreak Greater(\allowbreak Std(\allowbreak \$returns, 12)\allowbreak , Mean(\allowbreak Std(\allowbreak \$returns, 12)\allowbreak , 48)\allowbreak )\allowbreak , Neg(\allowbreak CsRank(\allowbreak Delta(\allowbreak \$close, 3)\allowbreak )\allowbreak )\allowbreak , Neg(\allowbreak CsRank(\allowbreak Div(\allowbreak Sub(\allowbreak \$close, \$low)\allowbreak , Add(\allowbreak Sub(\allowbreak \$high, \$low)\allowbreak , 0.0001)\allowbreak )\allowbreak )\allowbreak )\allowbreak )\allowbreak }}
\frow{047}{\texttt{Neg(\allowbreak Mul(\allowbreak CsRank(\allowbreak TsRank(\allowbreak Delta(\allowbreak Log(\allowbreak Add(\allowbreak \$volume, 1)\allowbreak )\allowbreak , 3)\allowbreak , 6)\allowbreak )\allowbreak , CsRank(\allowbreak Div(\allowbreak Delta(\allowbreak \$close, 6)\allowbreak , \$close)\allowbreak )\allowbreak )\allowbreak )\allowbreak }}
\frow{048}{\texttt{TsRank(\allowbreak Sub(\allowbreak CsRank(\allowbreak Std(\allowbreak \$returns, 12)\allowbreak )\allowbreak , CsRank(\allowbreak Delta(\allowbreak \$close, 6)\allowbreak )\allowbreak )\allowbreak , 18)\allowbreak }}
\frow{049}{\texttt{IfElse(\allowbreak Greater(\allowbreak Skew(\allowbreak \$returns, 24)\allowbreak , 0.5)\allowbreak , CsRank(\allowbreak Delta(\allowbreak \$volume, 3)\allowbreak )\allowbreak , Neg(\allowbreak CsRank(\allowbreak Std(\allowbreak \$returns, 12)\allowbreak )\allowbreak )\allowbreak )\allowbreak }}
\frow{050}{\texttt{IfElse(\allowbreak Greater(\allowbreak Skew(\allowbreak \$returns, 24)\allowbreak , 0.5)\allowbreak , Neg(\allowbreak CsRank(\allowbreak Div(\allowbreak Sub(\allowbreak \$close, \$low)\allowbreak , Add(\allowbreak Sub(\allowbreak \$high, \$low)\allowbreak , 1e-6)\allowbreak )\allowbreak )\allowbreak )\allowbreak , Neg(\allowbreak CsRank(\allowbreak Delta(\allowbreak \$close, 6)\allowbreak )\allowbreak )\allowbreak )\allowbreak }}
\frow{051}{\texttt{IfElse(\allowbreak Less(\allowbreak Skew(\allowbreak \$returns, 24)\allowbreak , -0.5)\allowbreak , CsRank(\allowbreak Delta(\allowbreak \$close, 5)\allowbreak )\allowbreak , Neg(\allowbreak CsRank(\allowbreak Std(\allowbreak \$returns, 12)\allowbreak )\allowbreak )\allowbreak )\allowbreak }}
\frow{052}{\texttt{IfElse(\allowbreak Less(\allowbreak Skew(\allowbreak \$returns, 24)\allowbreak , -0.5)\allowbreak , Neg(\allowbreak TsRank(\allowbreak \$returns, 12)\allowbreak )\allowbreak , Neg(\allowbreak CsRank(\allowbreak Delta(\allowbreak \$close, 6)\allowbreak )\allowbreak )\allowbreak )\allowbreak }}
\frow{053}{\texttt{Neg(\allowbreak CsRank(\allowbreak TsArgMax(\allowbreak SignedPower(\allowbreak IfElse(\allowbreak Less(\allowbreak \$returns, 0)\allowbreak , Std(\allowbreak \$returns, 20)\allowbreak , \$close)\allowbreak , 2)\allowbreak , 5)\allowbreak )\allowbreak )\allowbreak }}
\frow{054}{\texttt{IfElse(\allowbreak Greater(\allowbreak \$amt, Mean(\allowbreak \$amt, 24)\allowbreak )\allowbreak , CsRank(\allowbreak Delta(\allowbreak \$close, 3)\allowbreak )\allowbreak , Neg(\allowbreak CsRank(\allowbreak Div(\allowbreak Sub(\allowbreak \$close, \$vwap)\allowbreak , \$vwap)\allowbreak )\allowbreak )\allowbreak )\allowbreak }}
\frow{055}{\texttt{IfElse(\allowbreak Greater(\allowbreak Corr(\allowbreak \$close, \$volume, 12)\allowbreak , 0.5)\allowbreak , Neg(\allowbreak CsRank(\allowbreak \$returns)\allowbreak )\allowbreak , CsRank(\allowbreak Delta(\allowbreak \$volume, 3)\allowbreak )\allowbreak )\allowbreak }}
\frow{056}{\texttt{IfElse(\allowbreak Greater(\allowbreak Kurt(\allowbreak \$returns, 24)\allowbreak , 3.0)\allowbreak , Neg(\allowbreak CsRank(\allowbreak Div(\allowbreak \$amt, Add(\allowbreak \$volume, 1e-6)\allowbreak )\allowbreak )\allowbreak )\allowbreak , Neg(\allowbreak CsRank(\allowbreak Delta(\allowbreak \$close, 3)\allowbreak )\allowbreak )\allowbreak )\allowbreak }}
\frow{057}{\texttt{Neg(\allowbreak Sub(\allowbreak TsRank(\allowbreak Delta(\allowbreak \$close, 6)\allowbreak , 24)\allowbreak , TsRank(\allowbreak Delta(\allowbreak \$volume, 6)\allowbreak , 24)\allowbreak )\allowbreak )\allowbreak }}
\frow{058}{\texttt{IfElse(\allowbreak Greater(\allowbreak \$amt, Mean(\allowbreak \$amt, 48)\allowbreak )\allowbreak , Neg(\allowbreak CsRank(\allowbreak Div(\allowbreak \$amt, Add(\allowbreak \$volume, 1e-6)\allowbreak )\allowbreak )\allowbreak )\allowbreak , Neg(\allowbreak TsRank(\allowbreak \$returns, 12)\allowbreak )\allowbreak )\allowbreak }}
\frow{059}{\texttt{IfElse(\allowbreak Greater(\allowbreak Std(\allowbreak \$returns, 12)\allowbreak , Mean(\allowbreak Std(\allowbreak \$returns, 12)\allowbreak , 24)\allowbreak )\allowbreak , Neg(\allowbreak CsRank(\allowbreak Skew(\allowbreak \$amt, 12)\allowbreak )\allowbreak )\allowbreak , Neg(\allowbreak CsRank(\allowbreak Delta(\allowbreak \$close, 3)\allowbreak )\allowbreak )\allowbreak )\allowbreak }}
\frow{060}{\texttt{IfElse(\allowbreak Greater(\allowbreak Div(\allowbreak \$amt, Mean(\allowbreak \$amt, 24)\allowbreak )\allowbreak , 1.2)\allowbreak , Sub(\allowbreak CsRank(\allowbreak \$close)\allowbreak , CsRank(\allowbreak \$volume)\allowbreak )\allowbreak , Neg(\allowbreak CsRank(\allowbreak Delta(\allowbreak \$close, 3)\allowbreak )\allowbreak )\allowbreak )\allowbreak }}
\frow{061}{\texttt{Neg(\allowbreak Mul(\allowbreak TsRank(\allowbreak Delta(\allowbreak \$volume, 1)\allowbreak , 12)\allowbreak , TsRank(\allowbreak Delta(\allowbreak \$close, 1)\allowbreak , 12)\allowbreak )\allowbreak )\allowbreak }}
\frow{062}{\texttt{Neg(\allowbreak Div(\allowbreak CsRank(\allowbreak \$returns)\allowbreak , Add(\allowbreak Std(\allowbreak \$amt, 12)\allowbreak , 1e-6)\allowbreak )\allowbreak )\allowbreak }}
\frow{063}{\texttt{Neg(\allowbreak TsRank(\allowbreak Add(\allowbreak Add(\allowbreak CsRank(\allowbreak Delta(\allowbreak \$returns, 3)\allowbreak )\allowbreak , CsRank(\allowbreak Delta(\allowbreak \$amt, 3)\allowbreak )\allowbreak )\allowbreak , CsRank(\allowbreak Kurt(\allowbreak \$volume, 12)\allowbreak )\allowbreak )\allowbreak , 12)\allowbreak )\allowbreak }}
\frow{064}{\texttt{IfElse(\allowbreak Less(\allowbreak Corr(\allowbreak \$close, \$volume, 12)\allowbreak , -0.5)\allowbreak , CsRank(\allowbreak \$returns)\allowbreak , Neg(\allowbreak CsRank(\allowbreak Delta(\allowbreak \$close, 3)\allowbreak )\allowbreak )\allowbreak )\allowbreak }}
\frow{065}{\texttt{IfElse(\allowbreak Greater(\allowbreak Std(\allowbreak \$returns, 12)\allowbreak , Mean(\allowbreak Std(\allowbreak \$returns, 12)\allowbreak , 48)\allowbreak )\allowbreak , Neg(\allowbreak CsRank(\allowbreak Delta(\allowbreak \$amt, 6)\allowbreak )\allowbreak )\allowbreak , Neg(\allowbreak CsRank(\allowbreak Delta(\allowbreak \$close, 1)\allowbreak )\allowbreak )\allowbreak )\allowbreak }}
\frow{066}{\texttt{IfElse(\allowbreak Greater(\allowbreak Kurt(\allowbreak \$volume, 24)\allowbreak , 3)\allowbreak , Neg(\allowbreak CsRank(\allowbreak Delta(\allowbreak \$amt, 3)\allowbreak )\allowbreak )\allowbreak , Neg(\allowbreak CsRank(\allowbreak \$returns)\allowbreak )\allowbreak )\allowbreak }}
\frow{067}{\texttt{Neg(\allowbreak Mul(\allowbreak CsRank(\allowbreak Div(\allowbreak Sub(\allowbreak \$low, \$close)\allowbreak , Add(\allowbreak Sub(\allowbreak \$low, \$high)\allowbreak , 1e-6)\allowbreak )\allowbreak )\allowbreak , CsRank(\allowbreak Delta(\allowbreak \$amt, 6)\allowbreak )\allowbreak )\allowbreak )\allowbreak }}
\frow{068}{\texttt{Neg(\allowbreak IfElse(\allowbreak Less(\allowbreak Skew(\allowbreak \$returns, 24)\allowbreak , -0.5)\allowbreak , Neg(\allowbreak Sub(\allowbreak TsRank(\allowbreak \$close, 12)\allowbreak , TsRank(\allowbreak \$volume, 12)\allowbreak )\allowbreak )\allowbreak , CsRank(\allowbreak Delta(\allowbreak \$returns, 3)\allowbreak )\allowbreak )\allowbreak )\allowbreak }}
\frow{069}{\texttt{Neg(\allowbreak IfElse(\allowbreak Greater(\allowbreak Kurt(\allowbreak \$returns, 12)\allowbreak , 3.0)\allowbreak , Neg(\allowbreak CsRank(\allowbreak Std(\allowbreak \$returns, 6)\allowbreak )\allowbreak )\allowbreak , CsRank(\allowbreak Delta(\allowbreak \$returns, 3)\allowbreak )\allowbreak )\allowbreak )\allowbreak }}
\frow{070}{\texttt{Neg(\allowbreak Mul(\allowbreak CsRank(\allowbreak Div(\allowbreak Sub(\allowbreak \$close, \$low)\allowbreak , Add(\allowbreak Sub(\allowbreak \$high, \$low)\allowbreak , 0.0001)\allowbreak )\allowbreak )\allowbreak , TsRank(\allowbreak Std(\allowbreak \$returns, 24)\allowbreak , 24)\allowbreak )\allowbreak )\allowbreak }}
\frow{071}{\texttt{Mul(\allowbreak CsRank(\allowbreak Div(\allowbreak Sub(\allowbreak \$high, \$close)\allowbreak , Add(\allowbreak Sub(\allowbreak \$high, \$low)\allowbreak , 1e-6)\allowbreak )\allowbreak )\allowbreak , TsRank(\allowbreak Std(\allowbreak \$volume, 12)\allowbreak , 12)\allowbreak )\allowbreak }}
\frow{072}{\texttt{IfElse(\allowbreak Greater(\allowbreak Kurt(\allowbreak \$volume, 24)\allowbreak , 3.0)\allowbreak , Neg(\allowbreak Corr(\allowbreak \$close, \$volume, 12)\allowbreak )\allowbreak , Neg(\allowbreak TsRank(\allowbreak \$returns, 24)\allowbreak )\allowbreak )\allowbreak }}
\frow{073}{\texttt{IfElse(\allowbreak Greater(\allowbreak \$amt, EMA(\allowbreak \$amt, 12)\allowbreak )\allowbreak , Neg(\allowbreak TsRank(\allowbreak Delta(\allowbreak \$volume, 3)\allowbreak , 12)\allowbreak )\allowbreak , Neg(\allowbreak CsRank(\allowbreak \$returns)\allowbreak )\allowbreak )\allowbreak }}
\frow{074}{\texttt{Neg(\allowbreak Mul(\allowbreak CsRank(\allowbreak Div(\allowbreak Sub(\allowbreak \$close, \$low)\allowbreak , Add(\allowbreak Sub(\allowbreak \$high, \$low)\allowbreak , 1e-6)\allowbreak )\allowbreak )\allowbreak , TsRank(\allowbreak Skew(\allowbreak \$volume, 24)\allowbreak , 12)\allowbreak )\allowbreak )\allowbreak }}
\frow{075}{\texttt{Neg(\allowbreak Mul(\allowbreak CsRank(\allowbreak Div(\allowbreak \$returns, \$amt)\allowbreak )\allowbreak , TsRank(\allowbreak Std(\allowbreak \$volume, 20)\allowbreak , 20)\allowbreak )\allowbreak )\allowbreak }}
\frow{076}{\texttt{Neg(\allowbreak IfElse(\allowbreak Greater(\allowbreak Div(\allowbreak \$amt, Mean(\allowbreak \$amt, 24)\allowbreak )\allowbreak , 2.0)\allowbreak , Neg(\allowbreak Delta(\allowbreak \$close, 1)\allowbreak )\allowbreak , SMA(\allowbreak \$returns, 6)\allowbreak )\allowbreak )\allowbreak }}
\frow{077}{\texttt{Neg(\allowbreak Mul(\allowbreak TsRank(\allowbreak Std(\allowbreak Div(\allowbreak \$volume, Mean(\allowbreak \$volume, 24)\allowbreak )\allowbreak , 24)\allowbreak , 24)\allowbreak , TsRank(\allowbreak \$returns, 12)\allowbreak )\allowbreak )\allowbreak }}
\frow{078}{\texttt{Neg(\allowbreak Mul(\allowbreak CsRank(\allowbreak Div(\allowbreak \$returns, \$amt)\allowbreak )\allowbreak , TsRank(\allowbreak Kurt(\allowbreak \$volume, 20)\allowbreak , 20)\allowbreak )\allowbreak )\allowbreak }}
\frow{079}{\texttt{IfElse(\allowbreak Greater(\allowbreak Std(\allowbreak \$returns, 12)\allowbreak , EMA(\allowbreak Std(\allowbreak \$returns, 12)\allowbreak , 48)\allowbreak )\allowbreak , Neg(\allowbreak Div(\allowbreak Sub(\allowbreak \$close, \$low)\allowbreak , Add(\allowbreak Sub(\allowbreak \$high, \$low)\allowbreak , 1e-6)\allowbreak )\allowbreak )\allowbreak , Div(\allowbreak Sub(\allowbreak \$high, \$close)\allowbreak , Add(\allowbreak Sub(\allowbreak \$high, \$low)\allowbreak , 1e-6)\allowbreak )\allowbreak )\allowbreak }}
\frow{080}{\texttt{IfElse(\allowbreak Greater(\allowbreak Rsquare(\allowbreak \$close, 24)\allowbreak , 0.7)\allowbreak , Neg(\allowbreak CsRank(\allowbreak Slope(\allowbreak \$close, 24)\allowbreak )\allowbreak )\allowbreak , Neg(\allowbreak CsRank(\allowbreak Resi(\allowbreak \$close, 12)\allowbreak )\allowbreak )\allowbreak )\allowbreak }}
\frow{081}{\texttt{IfElse(\allowbreak And(\allowbreak Greater(\allowbreak Rsquare(\allowbreak \$close, 24)\allowbreak , 0.6)\allowbreak , Greater(\allowbreak Std(\allowbreak \$returns, 12)\allowbreak , Mean(\allowbreak Std(\allowbreak \$returns, 12)\allowbreak , 48)\allowbreak )\allowbreak )\allowbreak , Slope(\allowbreak \$close, 12)\allowbreak , Neg(\allowbreak SMA(\allowbreak \$returns, 6)\allowbreak )\allowbreak )\allowbreak }}
\frow{082}{\texttt{IfElse(\allowbreak Or(\allowbreak Greater(\allowbreak Div(\allowbreak Sub(\allowbreak \$high, Max2(\allowbreak \$open, \$close)\allowbreak )\allowbreak , Add(\allowbreak Sub(\allowbreak \$high, \$low)\allowbreak , 1e-6)\allowbreak )\allowbreak , 0.6)\allowbreak , Greater(\allowbreak Div(\allowbreak Sub(\allowbreak Min2(\allowbreak \$open, \$close)\allowbreak , \$low)\allowbreak , Add(\allowbreak Sub(\allowbreak \$high, \$low)\allowbreak , 1e-6)\allowbreak )\allowbreak , 0.6)\allowbreak )\allowbreak , Slope(\allowbreak \$close, 12)\allowbreak , Neg(\allowbreak SMA(\allowbreak \$returns, 6)\allowbreak )\allowbreak )\allowbreak }}
\frow{083}{\texttt{IfElse(\allowbreak And(\allowbreak Greater(\allowbreak Rsquare(\allowbreak \$close, 24)\allowbreak , 0.5)\allowbreak , Less(\allowbreak Std(\allowbreak \$returns, 12)\allowbreak , Mean(\allowbreak Std(\allowbreak \$returns, 12)\allowbreak , 48)\allowbreak )\allowbreak )\allowbreak , Slope(\allowbreak \$close, 12)\allowbreak , Neg(\allowbreak \$returns)\allowbreak )\allowbreak }}
\frow{084}{\texttt{IfElse(\allowbreak Eq(\allowbreak Sign(\allowbreak Delta(\allowbreak Std(\allowbreak \$returns, 12)\allowbreak , 1)\allowbreak )\allowbreak , Sign(\allowbreak Delta(\allowbreak \$returns, 1)\allowbreak )\allowbreak )\allowbreak , Neg(\allowbreak Resi(\allowbreak \$close, 12)\allowbreak )\allowbreak , Neg(\allowbreak Slope(\allowbreak \$close, 24)\allowbreak )\allowbreak )\allowbreak }}
\frow{085}{\texttt{Neg(\allowbreak Mul(\allowbreak CsRank(\allowbreak Rsquare(\allowbreak \$close, 24)\allowbreak )\allowbreak , CsRank(\allowbreak Resi(\allowbreak \$close, 12)\allowbreak )\allowbreak )\allowbreak )\allowbreak }}
\frow{086}{\texttt{IfElse(\allowbreak Eq(\allowbreak Sign(\allowbreak Delta(\allowbreak Resi(\allowbreak \$close, 12)\allowbreak , 1)\allowbreak )\allowbreak , Sign(\allowbreak Resi(\allowbreak \$close, 12)\allowbreak )\allowbreak )\allowbreak , Neg(\allowbreak Resi(\allowbreak \$close, 6)\allowbreak )\allowbreak , Neg(\allowbreak Slope(\allowbreak \$close, 12)\allowbreak )\allowbreak )\allowbreak }}
\frow{087}{\texttt{IfElse(\allowbreak Greater(\allowbreak Kurt(\allowbreak \$returns, 24)\allowbreak , 3.0)\allowbreak , Neg(\allowbreak Resi(\allowbreak \$close, 6)\allowbreak )\allowbreak , Neg(\allowbreak Resi(\allowbreak \$close, 24)\allowbreak )\allowbreak )\allowbreak }}
\frow{088}{\texttt{Neg(\allowbreak Mul(\allowbreak CsRank(\allowbreak Div(\allowbreak \$returns, Add(\allowbreak \$amt, 1e-6)\allowbreak )\allowbreak )\allowbreak , TsRank(\allowbreak Skew(\allowbreak \$returns, 24)\allowbreak , 24)\allowbreak )\allowbreak )\allowbreak }}
\frow{089}{\texttt{IfElse(\allowbreak Or(\allowbreak Greater(\allowbreak \$volume, Mean(\allowbreak \$volume, 24)\allowbreak )\allowbreak , Greater(\allowbreak Abs(\allowbreak \$returns)\allowbreak , Std(\allowbreak \$returns, 24)\allowbreak )\allowbreak )\allowbreak , Neg(\allowbreak Resi(\allowbreak \$close, 6)\allowbreak )\allowbreak , Neg(\allowbreak Slope(\allowbreak \$close, 12)\allowbreak )\allowbreak )\allowbreak }}
\frow{090}{\texttt{IfElse(\allowbreak Greater(\allowbreak Rsquare(\allowbreak \$close, 24)\allowbreak , 0.75)\allowbreak , Neg(\allowbreak Slope(\allowbreak \$close, 12)\allowbreak )\allowbreak , Neg(\allowbreak TsRank(\allowbreak \$returns, 12)\allowbreak )\allowbreak )\allowbreak }}
\frow{091}{\texttt{Neg(\allowbreak Mul(\allowbreak TsRank(\allowbreak Div(\allowbreak \$returns, Add(\allowbreak \$amt, 1e-6)\allowbreak )\allowbreak , 24)\allowbreak , TsRank(\allowbreak Kurt(\allowbreak \$returns, 24)\allowbreak , 24)\allowbreak )\allowbreak )\allowbreak }}
\frow{092}{\texttt{Neg(\allowbreak CsRank(\allowbreak EMA(\allowbreak Div(\allowbreak \$returns, Add(\allowbreak \$amt, 1e-6)\allowbreak )\allowbreak , 6)\allowbreak )\allowbreak )\allowbreak }}
\frow{093}{\texttt{Neg(\allowbreak TsRank(\allowbreak Delta(\allowbreak Div(\allowbreak \$returns, Add(\allowbreak \$amt, 1e-6)\allowbreak )\allowbreak , 3)\allowbreak , 24)\allowbreak )\allowbreak }}
\frow{094}{\texttt{IfElse(\allowbreak And(\allowbreak Greater(\allowbreak Std(\allowbreak \$returns, 12)\allowbreak , Mean(\allowbreak Std(\allowbreak \$returns, 12)\allowbreak , 60)\allowbreak )\allowbreak , Less(\allowbreak Kurt(\allowbreak \$returns, 24)\allowbreak , 2)\allowbreak )\allowbreak , Neg(\allowbreak Delta(\allowbreak \$close, 3)\allowbreak )\allowbreak , Neg(\allowbreak TsRank(\allowbreak \$returns, 24)\allowbreak )\allowbreak )\allowbreak }}
\frow{095}{\texttt{IfElse(\allowbreak Or(\allowbreak Greater(\allowbreak Abs(\allowbreak Skew(\allowbreak \$returns, 24)\allowbreak )\allowbreak , 1.5)\allowbreak , Greater(\allowbreak Kurt(\allowbreak \$returns, 24)\allowbreak , 4.0)\allowbreak )\allowbreak , Neg(\allowbreak Resi(\allowbreak \$close, 6)\allowbreak )\allowbreak , Neg(\allowbreak TsRank(\allowbreak \$returns, 24)\allowbreak )\allowbreak )\allowbreak }}
\frow{096}{\texttt{Neg(\allowbreak Mul(\allowbreak TsRank(\allowbreak WMA(\allowbreak Div(\allowbreak \$returns, Add(\allowbreak \$amt, 1e-6)\allowbreak )\allowbreak , 6)\allowbreak , 24)\allowbreak , TsRank(\allowbreak Corr(\allowbreak \$close, \$volume, 24)\allowbreak , 24)\allowbreak )\allowbreak )\allowbreak }}
\frow{097}{\texttt{Neg(\allowbreak CsRank(\allowbreak Med(\allowbreak Div(\allowbreak \$returns, Add(\allowbreak \$amt, 1e-6)\allowbreak )\allowbreak , 6)\allowbreak )\allowbreak )\allowbreak }}
\frow{098}{\texttt{Neg(\allowbreak CsRank(\allowbreak Delta(\allowbreak WMA(\allowbreak Div(\allowbreak \$returns, Add(\allowbreak \$amt, 1e-6)\allowbreak )\allowbreak , 6)\allowbreak , 3)\allowbreak )\allowbreak )\allowbreak }}
\frow{099}{\texttt{Neg(\allowbreak Sub(\allowbreak EMA(\allowbreak Div(\allowbreak \$returns, Add(\allowbreak \$amt, 1e-6)\allowbreak )\allowbreak , 6)\allowbreak , EMA(\allowbreak Div(\allowbreak \$returns, Add(\allowbreak \$amt, 1e-6)\allowbreak )\allowbreak , 24)\allowbreak )\allowbreak )\allowbreak }}
\frow{100}{\texttt{Neg(\allowbreak Resi(\allowbreak Delta(\allowbreak Resi(\allowbreak \$close, 24)\allowbreak , 3)\allowbreak , 12)\allowbreak )\allowbreak }}
\frow{101}{\texttt{Neg(\allowbreak IfElse(\allowbreak Greater(\allowbreak Abs(\allowbreak Skew(\allowbreak \$returns, 24)\allowbreak )\allowbreak , 1.0)\allowbreak , TsRank(\allowbreak Div(\allowbreak Sub(\allowbreak \$returns, Med(\allowbreak \$returns, 24)\allowbreak )\allowbreak , Add(\allowbreak Std(\allowbreak \$returns, 24)\allowbreak , 1e-6)\allowbreak )\allowbreak , 24)\allowbreak , CsRank(\allowbreak Resi(\allowbreak \$close, 12)\allowbreak )\allowbreak )\allowbreak )\allowbreak }}
\frow{102}{\texttt{IfElse(\allowbreak Greater(\allowbreak Skew(\allowbreak \$high, 24)\allowbreak , 0)\allowbreak , TsRank(\allowbreak Log(\allowbreak Div(\allowbreak \$high, Add(\allowbreak \$close, 1e-6)\allowbreak )\allowbreak )\allowbreak , 24)\allowbreak , CsRank(\allowbreak Resi(\allowbreak \$low, 24)\allowbreak )\allowbreak )\allowbreak }}
\frow{103}{\texttt{Neg(\allowbreak IfElse(\allowbreak Greater(\allowbreak Skew(\allowbreak \$open, 24)\allowbreak , 0)\allowbreak , TsRank(\allowbreak Resi(\allowbreak \$open, 24)\allowbreak , 24)\allowbreak , CsRank(\allowbreak SignedPower(\allowbreak \$returns, 0.5)\allowbreak )\allowbreak )\allowbreak )\allowbreak }}
\frow{104}{\texttt{Neg(\allowbreak IfElse(\allowbreak Greater(\allowbreak Med(\allowbreak \$returns, 24)\allowbreak , 0)\allowbreak , TsRank(\allowbreak Log(\allowbreak Div(\allowbreak \$close, Add(\allowbreak \$open, 1e-6)\allowbreak )\allowbreak )\allowbreak , 24)\allowbreak , CsRank(\allowbreak Sign(\allowbreak Resi(\allowbreak \$close, 24)\allowbreak )\allowbreak )\allowbreak )\allowbreak )\allowbreak }}
\frow{105}{\texttt{Neg(\allowbreak IfElse(\allowbreak Greater(\allowbreak Skew(\allowbreak \$returns,24)\allowbreak ,0.4)\allowbreak , CsRank(\allowbreak Sign(\allowbreak Resi(\allowbreak \$open,24)\allowbreak )\allowbreak )\allowbreak , TsRank(\allowbreak SignedPower(\allowbreak \$returns,0.6)\allowbreak ,24)\allowbreak )\allowbreak )\allowbreak }}
\frow{106}{\texttt{Neg(\allowbreak IfElse(\allowbreak Greater(\allowbreak Std(\allowbreak \$volume,24)\allowbreak ,Mean(\allowbreak Std(\allowbreak \$volume,24)\allowbreak ,48)\allowbreak )\allowbreak , CsRank(\allowbreak Div(\allowbreak \$amt, Add(\allowbreak \$volume,1e-6)\allowbreak )\allowbreak )\allowbreak , TsRank(\allowbreak Div(\allowbreak \$returns, Add(\allowbreak Std(\allowbreak \$returns,24)\allowbreak ,1e-6)\allowbreak )\allowbreak ,24)\allowbreak )\allowbreak )\allowbreak }}
\frow{107}{\texttt{IfElse(\allowbreak Greater(\allowbreak Std(\allowbreak \$returns, 12)\allowbreak , Med(\allowbreak Std(\allowbreak \$returns, 12)\allowbreak , 48)\allowbreak )\allowbreak , Mul(\allowbreak CsRank(\allowbreak Neg(\allowbreak Div(\allowbreak Sub(\allowbreak \$close, \$vwap)\allowbreak , \$vwap)\allowbreak )\allowbreak )\allowbreak , CsRank(\allowbreak Div(\allowbreak \$volume, EMA(\allowbreak \$volume, 12)\allowbreak )\allowbreak )\allowbreak )\allowbreak , Neg(\allowbreak TsRank(\allowbreak Div(\allowbreak Sub(\allowbreak \$close, \$vwap)\allowbreak , \$vwap)\allowbreak , 24)\allowbreak )\allowbreak )\allowbreak }}
\frow{108}{\texttt{Neg(\allowbreak IfElse(\allowbreak Less(\allowbreak Skew(\allowbreak \$returns, 24)\allowbreak , -0.3)\allowbreak , Mul(\allowbreak CsRank(\allowbreak Neg(\allowbreak Div(\allowbreak Sub(\allowbreak \$close, \$vwap)\allowbreak , \$vwap)\allowbreak )\allowbreak )\allowbreak , CsRank(\allowbreak Div(\allowbreak \$volume, EMA(\allowbreak \$volume, 12)\allowbreak )\allowbreak )\allowbreak )\allowbreak , TsRank(\allowbreak Div(\allowbreak Sub(\allowbreak \$close, \$vwap)\allowbreak , \$vwap)\allowbreak , 12)\allowbreak )\allowbreak )\allowbreak }}
\frow{109}{\texttt{IfElse(\allowbreak Greater(\allowbreak Rsquare(\allowbreak \$close, 24)\allowbreak , 0.6)\allowbreak , Neg(\allowbreak CsRank(\allowbreak Slope(\allowbreak \$close, 24)\allowbreak )\allowbreak )\allowbreak , Mul(\allowbreak CsRank(\allowbreak Neg(\allowbreak Div(\allowbreak Sub(\allowbreak \$close, \$vwap)\allowbreak , \$vwap)\allowbreak )\allowbreak )\allowbreak , CsRank(\allowbreak Div(\allowbreak \$volume, EMA(\allowbreak \$volume, 12)\allowbreak )\allowbreak )\allowbreak )\allowbreak )\allowbreak }}
\frow{110}{\texttt{Neg(\allowbreak Mul(\allowbreak CsRank(\allowbreak Div(\allowbreak Sub(\allowbreak \$close, \$low)\allowbreak , Add(\allowbreak Sub(\allowbreak \$high, \$low)\allowbreak , 1e-6)\allowbreak )\allowbreak )\allowbreak , CsRank(\allowbreak Mul(\allowbreak Skew(\allowbreak \$returns, 24)\allowbreak , Slope(\allowbreak \$close, 12)\allowbreak )\allowbreak )\allowbreak )\allowbreak )\allowbreak }}
\par\vskip1pt\noindent\rule{\columnwidth}{0.6pt}
\endgroup

\end{document}